\documentclass[twocolumn,showpacs,preprintnumbers,amsmath,amssymb]{revtex4}

\usepackage{graphicx}
\usepackage{dcolumn}
\usepackage{bm}
\usepackage{multirow}

\usepackage{color}

\begin{document}

\title{Polymer adhesion: first-principles
  calculations of the adsorption of organic molecules onto Si
  surfaces.  } 
\author{Karen Johnston}
\affiliation{Laboratory of Physics, Helsinki University of
  Technology, P.O. Box 1100, 02015, Finland}
\author{Risto M. Nieminen}
\affiliation{Laboratory of Physics, Helsinki University of
  Technology, P.O. Box 1100, 02015, Finland}

\begin{abstract}
The structures and energetics of organic molecules adsorbed onto
clean and H-passivated Si(001)-(2$\times$1) surfaces have been
calculated using density functional theory.  
For benzene adsorbed on the clean Si surface the tight-bridge
structure was found to be stable and the butterfly structure
metastable.   
Both carbonic acid H$_2$CO$_3$ and propane C$_3$H$_8$ dissociate
on contact with the surface.  
Passivation of the Si surface with H-atoms has a dramatic effect
on the surface properties.  The passivated surface is very inert
and the binding energy of all the molecules is very weak.  
\end{abstract}
\maketitle

The adsorption of organic molecules on semi-conductor surfaces is
of increasing importance to industry due to interest in the 
development of organic optoelectronic devices,
micro/molecular-scale electronics and biofunctionality \cite{Lopinski2000a,Besley2006a,Quek2006a}.  An
additional motivation for studying these systems is to understand
the nature of adhesion between plastics and metal or ceramic
surfaces.  Many of these materials do not adhere well and the
main focus of this research is to obtain a better understanding
of the structure and bonding at the polymer-surface interface.   

\begin{figure}[ht!]
\begin{center}
\includegraphics[height=2cm]{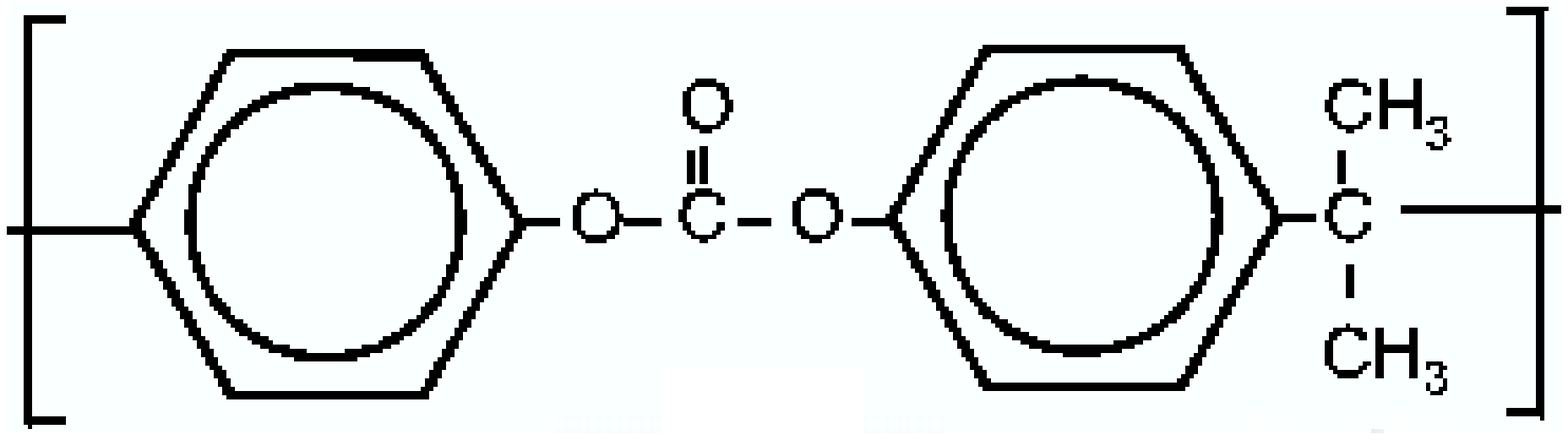}
\includegraphics[height=1.8cm]{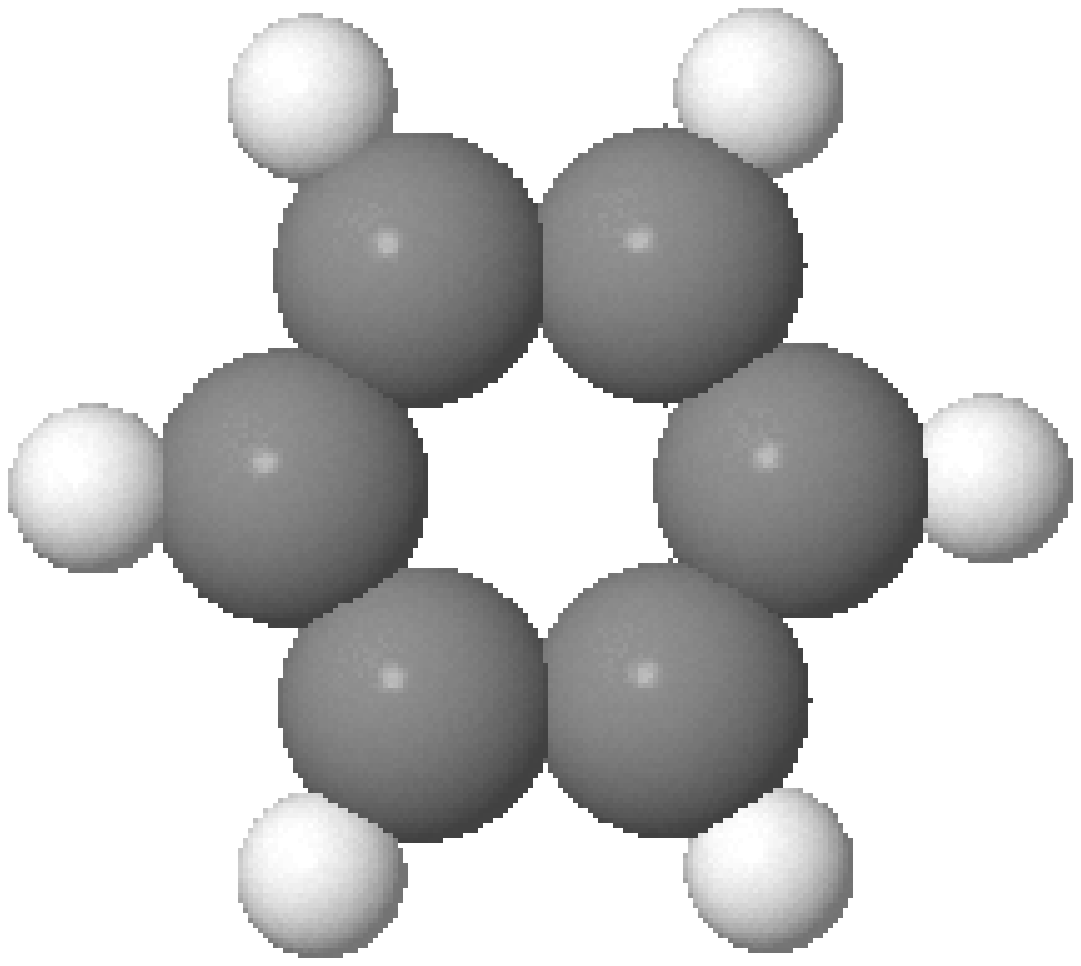}
\includegraphics[height=1.1cm]{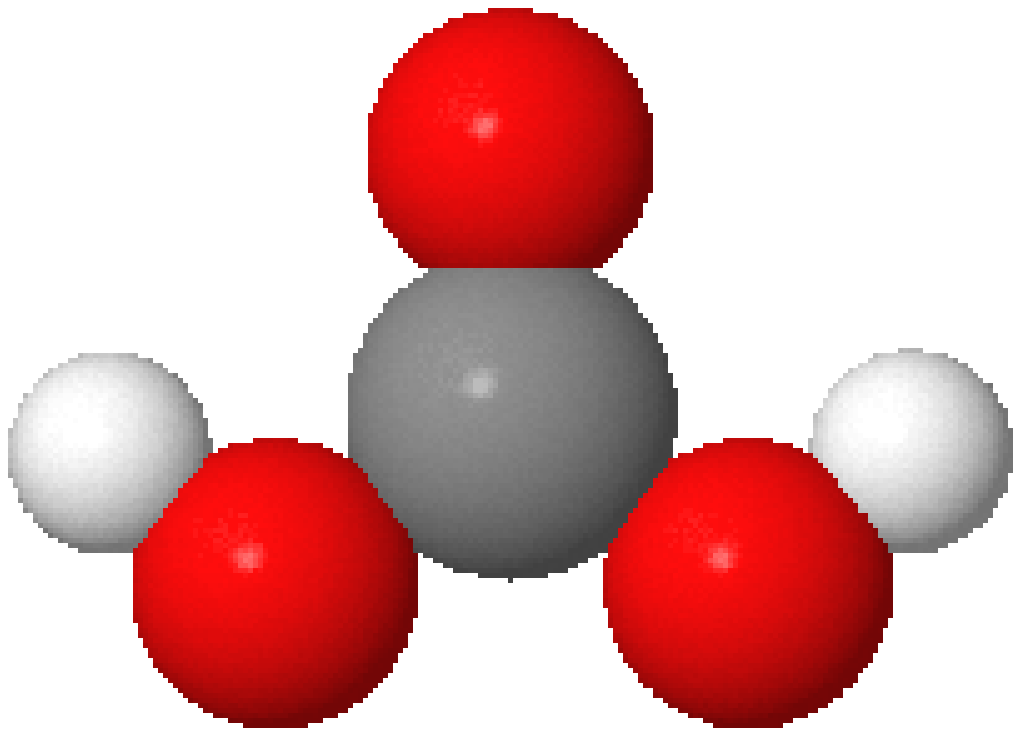}
\includegraphics[height=1.8cm]{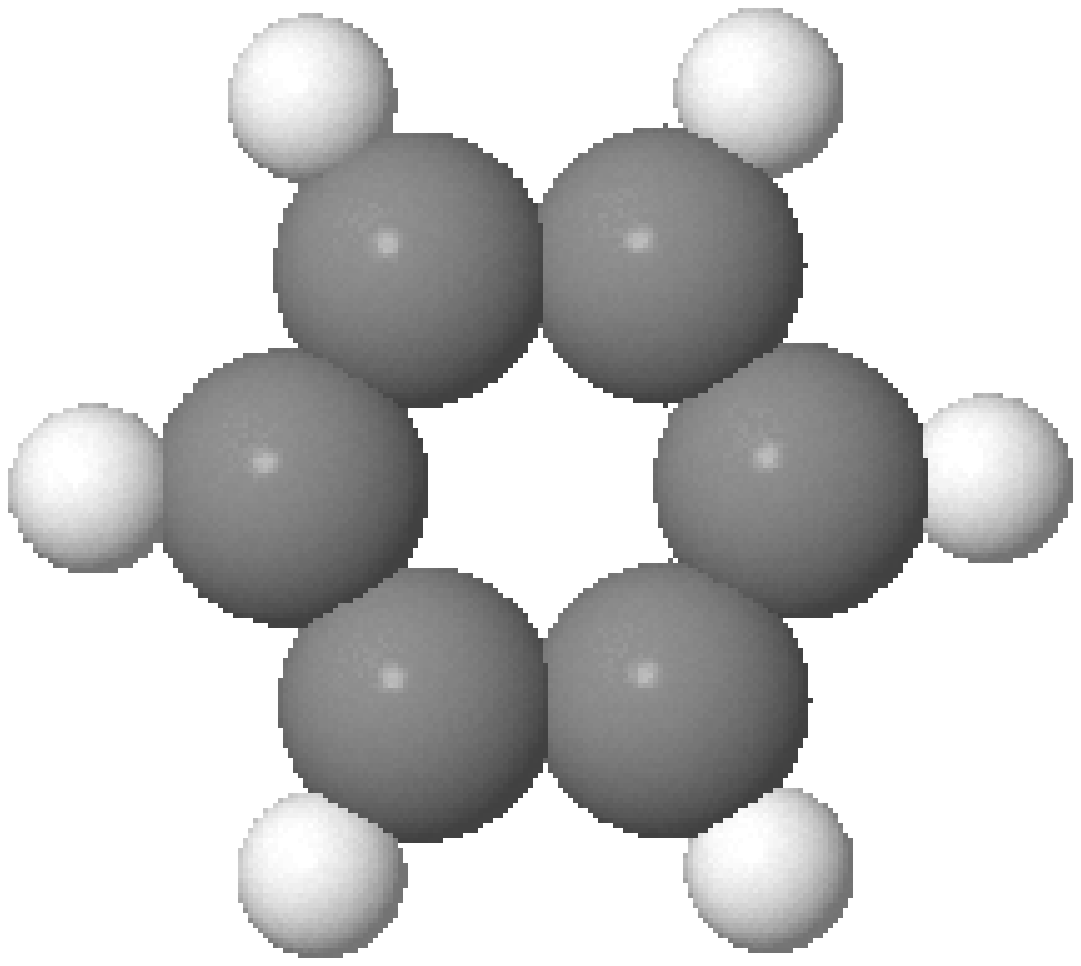}
\includegraphics[height=1.8cm]{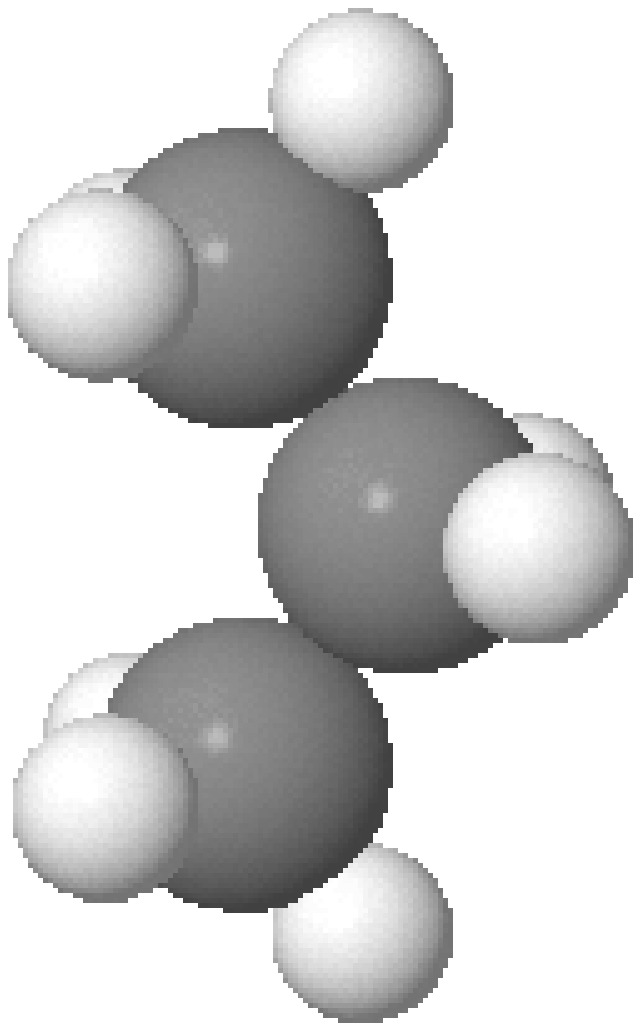}
\caption{\label{fig:BPA-PC} (Color online) The monomer
bisphenol-A-polycarbonate (BPA-PC) and its analogous molecules
benzene (C$_6$H$_6$), carbonic acid (H$_2$CO$_3$), another
benzene and propane (C$_3$H$_8$).  }
\end{center}
\end{figure}

While oxide/ceramic surfaces are primarily of interest, we have
chosen to first study the simpler Si(001) surface, which is
easier to simulate and will provide the initial insight into the
nature of the bonding at surfaces.  This will provide the basis
for future work involving the more complex silica surface.   
The plastic of interest is mainly composed of the polymer
bisphenol-A-polycarbonate (BPA-PC).  The repeat unit, or monomer,
is shown in Figure~\ref{fig:BPA-PC}.  
This polymer has been extensively studied with both
first-principles methods
\cite{Montanari1998a,Montanari1998b,Montanari1999a} and
coarse-grained molecular dynamics simulations \cite{Abrams2003b}.  

To understand the nature of adhesion it is necessary to consider
processes which occur at different time, length and temperature
scales and therefore a multiscale study is desirable.
First-principles calculations provide details in the microscopic
regime but they are too computationally demanding to be able to
treat the entire polymer molecule on a surface.  It is,
therefore, necessary to divide the monomer into its analogous
molecules (benzene, propane and carbonic acid as shown in
Figure~\ref{fig:BPA-PC}) and study their individual adsorption
on the surface.    
The results of these first-principles calculations could then
be used to build potentials for coarse-grained simulations, so
these calculations are the first step towards a multiscale study. 

For low density systems, such as molecular adsorption, van der
Waals forces can play a significant role.  Recently a method for
including the van der Waals energy within density functional theory
(DFT) was suggested \cite{Dion2004a,Dion2005a} and has been shown
to determine the correct binding distances and energies for benzene
dimers \cite{Puzder2006a,Thonhauser2006a} and for benzene and
naphthalene adsorbed on graphite
\cite{Chakarova2006a,Chakarova2006b}.  
However, it is known that benzene binds strongly to Si and the
additional binding due to van der Waals forces will be small
compared to the chemical binding energies.  Thus, the effect of
van der Waals forces has not been taken into account in this
paper.  

The paper is organised as follows:  Section~\ref{sec:surfaces}
describes the two Si surfaces under consideration and gives the
technical details of the calculations.  Sections~\ref{sec:clean-Si}
and \ref{sec:H-Si} describe the adsorption of benzene and the other
organic molecules on the clean and H-passivated Si dimer
surfaces, denoted Si(001)-(2$\times$1) and
Si(001):H-(2$\times$1), respectively.  The discussion and
conclusions appear in Section~\ref{sec:summary}.  

\section{Si(001)-(2$\times$1) and Si(001):H-(2$\times$1) surfaces}
\label{sec:surfaces}

\begin{figure}[th!]
\begin{center}
\includegraphics[height=2.0cm,angle=0]{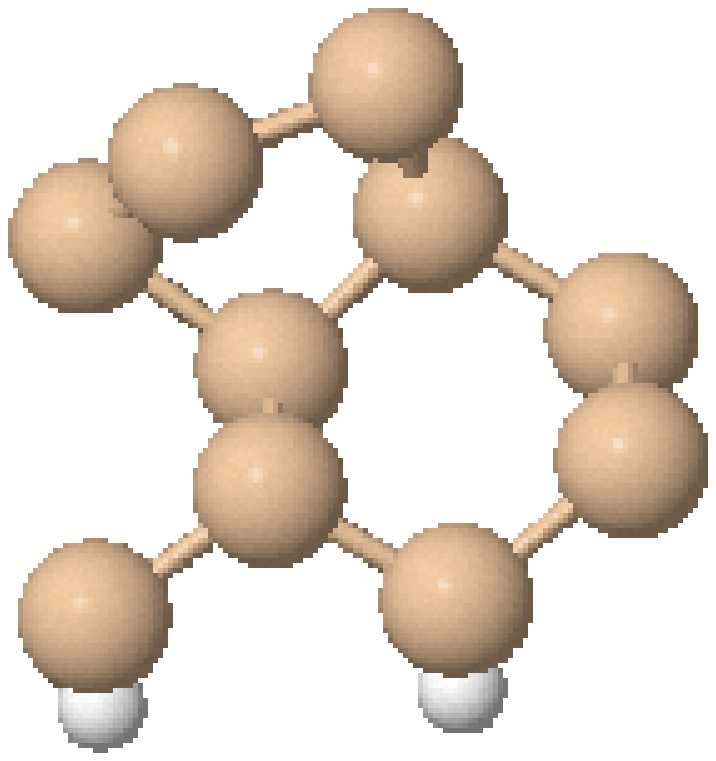}
\includegraphics[height=2.3cm,angle=0]{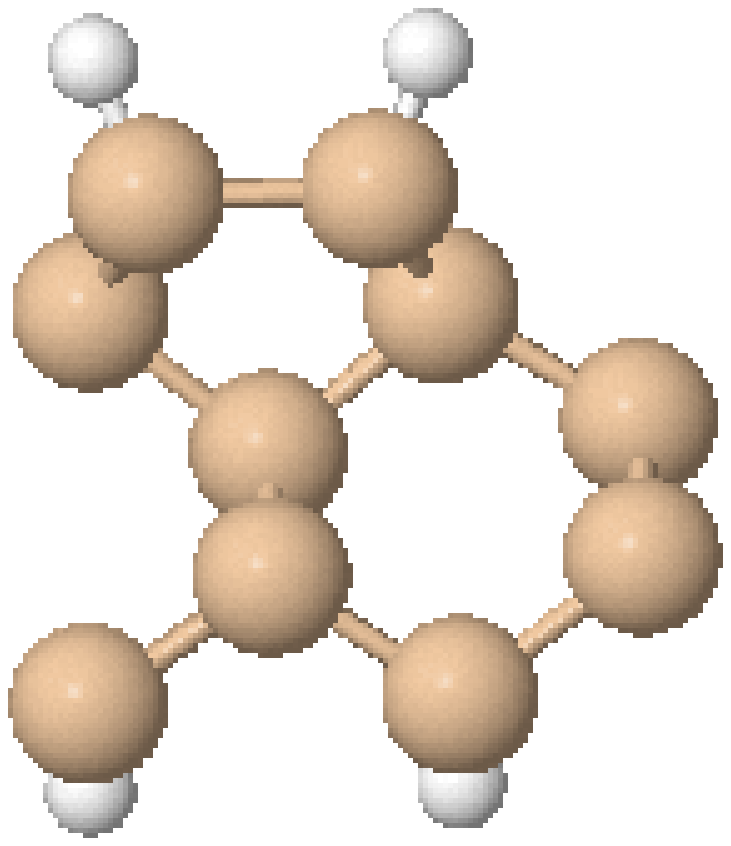}
\caption{\label{fig:Si-2x1} (Color online) The clean
Si(001)-(2$\times$1) dimer reconstruction (left) and H-passivated
Si(001):H-(2$\times$1) surface (right) viewed along the [110]
direction. } 
\end{center}
\end{figure}

At room temperature the Si(001) surface has a (2$\times$1)
structure, which consists of the formation of buckled
dimers, with one atom being drawn towards the surface closer to
its three nearest neighbours, and the other being pushed away
from its neighbours, as shown in Figure~\ref{fig:Si-2x1}.  
As the clean Si(001)-(2$\times$1) surface is rather reactive,
adsorbed atoms or molecules, such as H or H$_2$O, are likely to
be present and, therefore, we have also considered adsorption on
the H-passivated Si(001):H-(2$\times$1) surface.  The
H-(2$\times$1) surface is passivated by one H atom per Si atom,
as shown in Fig.~\ref{fig:Si-2x1}.  

First-principles calculations were performed with the Vienna {\it
Ab initio} Simulation Package (VASP)
\cite{Kresse1996a,Kresse1996b}, which implements density 
functional theory.  The PW91 flavour
\cite{Perdew1991a,Perdew1996b} of the generalised-gradient 
approximation (GGA) and projector augmented-wave potentials (PAW)
\cite{Blochl1994a} were used.   
The plane-wave energy cutoff was 400~eV, which corresponds to an
error in the total energy of $\approx$3~meV per Si atom, and the
Brillouin zone mesh was equivalent to a 4$\times$4$\times$1
Monkhorst-Pack $k$-point sampling for a 0.5~ML supercell with
dimensions (1 1 0) (-1 1 0) (0 0 4).  
The calculated lattice constant of bulk Si is 5.47~{\AA}.  

For the slab calculations we used a 5-atomic layer Si slab and
$\approx$15~{\AA} of vacuum.  All relaxations were considered
complete when the forces were less than 10~meV{\AA}$^{-1}$.  
To save computational time only adsorption on one side of the
slab was considered.  
When the slab is asymmetric and periodic boundary conditions are
used, as is the case in these calculations, it must be ensured
that there is no dipole-dipole interaction between the slabs.
The energy of the dipole-dipole interaction is inversely
proportional to the distance between the slabs so one way to
avoid this interaction is to have a large vacuum region between
slabs.  To check this the vacuum was increased from
$\approx$15~{\AA} to $\approx$25~{\AA} and it was found that the
adsorption energies did not change.  This indicates that the
dipole-dipole interaction is negligible.  
For both the clean and H-passivated
slabs the bottom layer of the slab was fixed in the bulk Si
positions and passivated with two H atoms per Si atom.   
The Si(001)-(2$\times$1) surface structure of a 9-atomic layer
slab was also calculated and the differences between the 9 and
5-atomic layer slabs were small, i.e. the dimer bond lengths are
2.31~{\AA} and 2.30~{\AA} and the angles are 18.11$^{\rm o}$ and
18.02$^{\rm o}$,
respectively.  Increasing the plane-wave cut-off energy to 600~eV
does not change these values.  To check that the chemisorption
energies are converged we also calculated the benzene adsorption
energies on an 9-atomic layer slab (see
subsection~\ref{sec:clean-Si-benzene}).  
The results presented in the paper use H-passivated 5-atomic
layer slabs and a cut-off of 400~eV unless otherwise indicated.
For the Si(001):H-(2$\times$1) surface the H-passivation caused the
dimer to flatten and the bondlength to increase to 2.42~{\AA}, in
agreement with Zanella {\it et al} \cite{Zanella2006a}.

\section{Adsorption on Si(001)-(2$\times$1)}
\label{sec:clean-Si}

\subsection{Benzene, C$_6$H$_6$}
\label{sec:clean-Si-benzene}

There are several publications which report the structural
geometry and energetics of benzene adsorbed on
Si(001)-(2$\times$1).  Five different geometries have been
studied in the literature: two single dimer di-$\sigma$ bonded 
structures named ``tilted'' and ``butterfly'', and three double
dimer tetra-$\sigma$ bonded structures: ``tight bridge'',
``twisted bridge'' and the symmetric bridge (``pedestal'').  The
two most stable structures are agreed to be the butterfly and
tight-bridge structures but to date there is no conclusive
experimental or computational study that determines which
structure is the stable one and which is metastable.  Here we
make a comprehensive overview of the available data and
present results for benzene adsorption at low coverages.  

\subsubsection{Structural data}

\begin{figure}[ht!]
\begin{center}
\includegraphics[height=2.5cm]{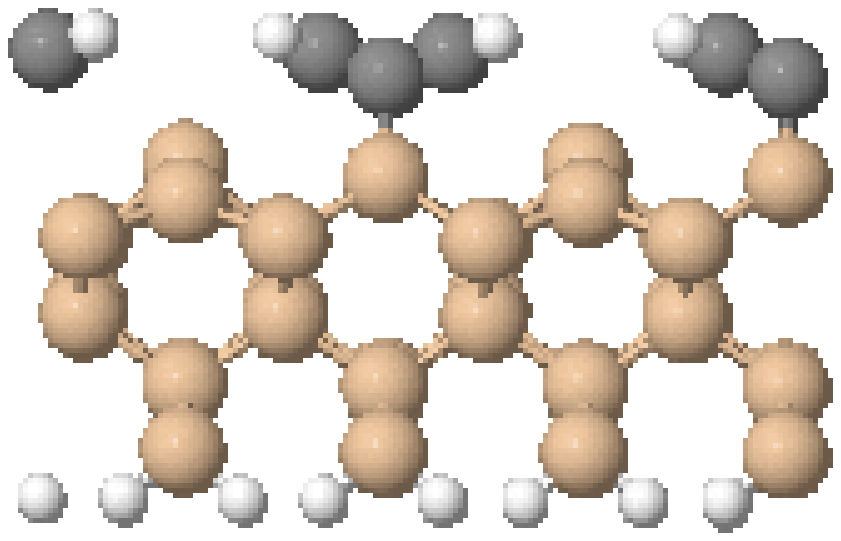}
\includegraphics[height=2.5cm]{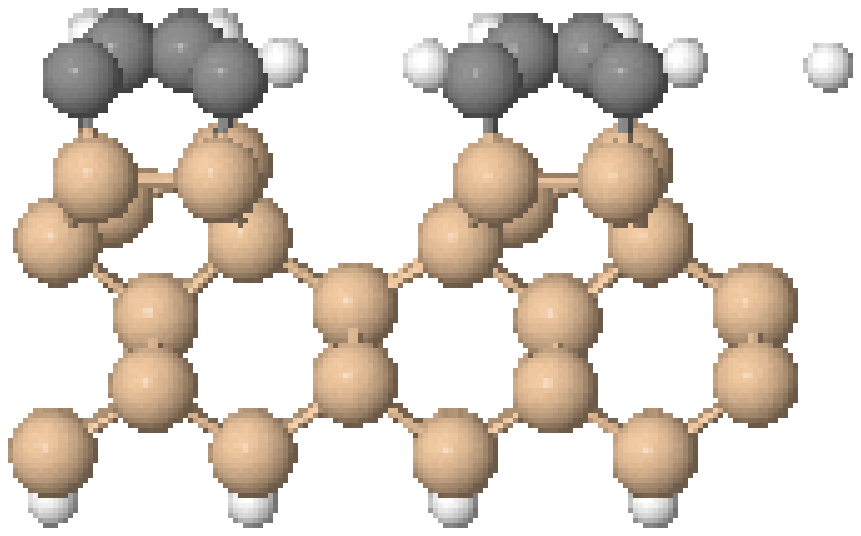}
\caption{\label{fig:butterfly} (Color online)
``Butterfly'' structure of C$_6$H$_6$ on Si(001)-(2$\times$1)
  viewed along [$\bar{1}10$] and [110] (dimer rows), respectively.  }
\end{center}
\end{figure}

\begin{figure}[ht!]
\begin{center}
\includegraphics[height=2.5cm]{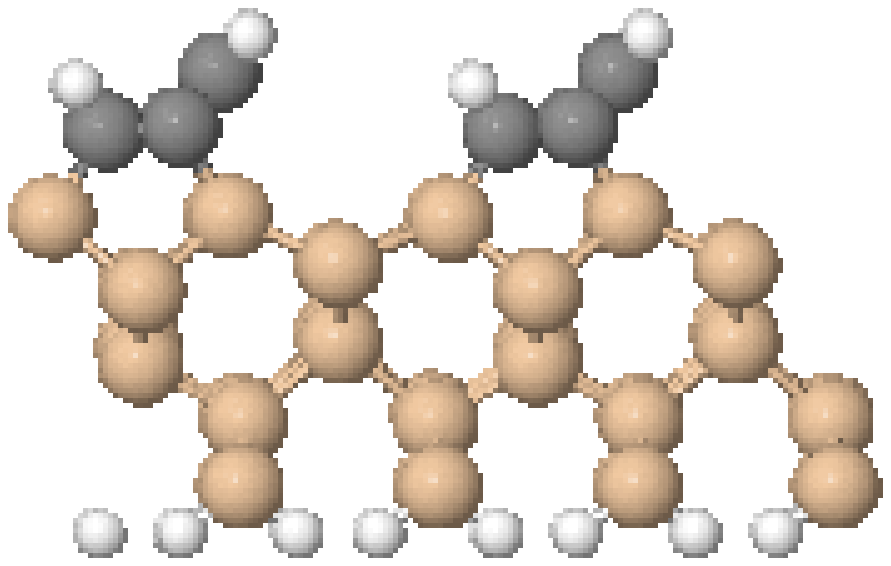}
\includegraphics[height=2.5cm]{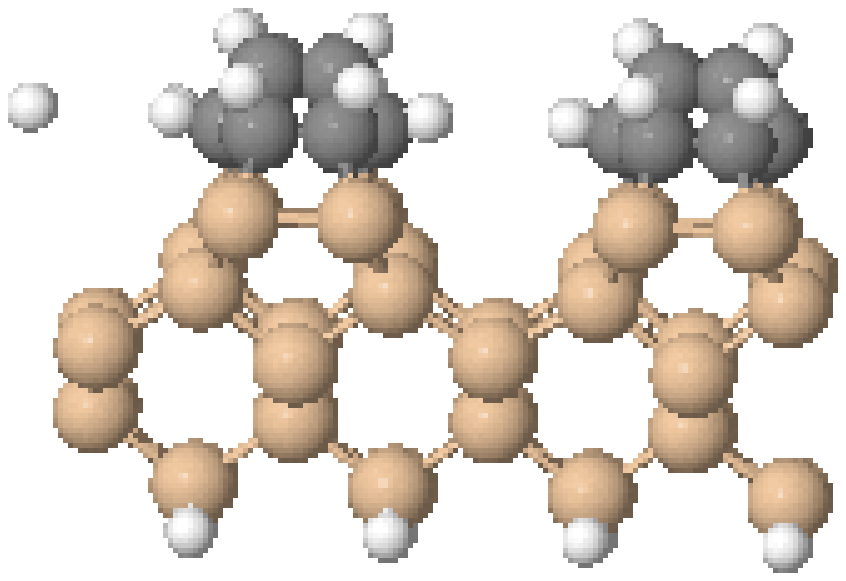}
\caption{\label{fig:tight-bridge} (Color online)
``Tight-bridge'' structure of C$_6$H$_6$ on Si(001)-(2$\times$1)
viewed along [$\bar{1}10$] and [110], respectively.}
\end{center}
\end{figure}

Figures~\ref{fig:butterfly} and \ref{fig:tight-bridge} show the
butterfly and tight-bridge structures, respectively.  The
butterfly structure is bonded to the two dangling bonds of a
single Si--Si dimer whereas the tight-bridge structure is bonded
to two Si--Si dimers.  These adsorption structures are
significantly distorted compared to the isolated benzene
molecule.  Another distinguishing feature is that the butterfly
structure has a symmetry plane along the Si-dimer.  

Our structural data for both geometries is in excellent
agreement with previous first-principles results (see
Tables~\ref{tab:butterfly} and \ref{tab:tight-bridge}).  In the
butterfly structure the benzene adopts a V-shape as shown in
Fig~\ref{fig:butterfly}.  The Si dimer bond length is 2.39{\AA},
which is in excellent agreement with Hofer {\it et al}
\cite{Hofer2001a}.   
\begin{table}[ht!]
\begin{center}
\begin{tabular}{|rrrrr|l|}\hline
 Si--Si 
&Si$_{\rm C}$--Si$_{\rm C}$
&Si$_{\rm C}$--C$_{\rm Si}$ 
&C--C$_{\rm Si}$
&C=C& Reference\\ \hline
2.34 &2.39 & 1.98 & 1.50      & 1.35 & Present\\
2.24 &2.34 & -    & 1.51      & 1.34 & \cite{Jung2005a}\\
-    &2.39 & 1.98 & 1.50      & 1.34 & \cite{Hofer2001a}\\
-    &-    & 1.89 & 1.47-1.49 & 1.35 & \cite{Wolkow1998a}\\
-    &2.46 & 1.97 & 1.51      & -    & \cite{Birkenheuer1998a}\\
\hline
\end{tabular}
\caption{\label{tab:butterfly} Bond lengths (\AA) for the butterfly
  geometry.  Si$_{\rm C}$ and C$_{\rm Si}$ denote Si atoms bonded
  to C and C atoms bonded to Si, respectively.} 
\end{center}
\end{table}
\begin{table}[ht!]
\begin{center}
\begin{tabular}{|rrrrr|l|}\hline
Si$_{\rm C}$--Si$_{\rm C}$
&Si$_{\rm C}$--C$_{\rm Si}$
&C$_{\rm Si}$--C$_{\rm Si}$
&C--C$_{\rm Si}$
&C=C
&Reference\\ \hline
2.35-2.38 & 1.99-2.00 & 1.57-1.58 & 1.50 & 1.35 & Present \\
     2.34 & 1.98-2.00 & 1.58-1.59 & 1.51 & 1.34 & \cite{Jung2005a}\\
     2.39 & 1.98-2.01 &      1.57 & 1.50 & 1.35 & \cite{Hofer2001a}\\
        - & 1.87-1.89 & 1.47-1.53 &    - & 1.35 & \cite{Wolkow1998a}\\
\hline
\end{tabular}
\caption{\label{tab:tight-bridge} Bond lengths (\AA) for
  tight-bridge geometry.  }
\end{center}
\end{table}

\subsubsection{Stability -- experimental evidence}

According to a scanning tunnelling microscopy (STM) study by
Lopinski {\it et al} \cite{Lopinski1998a}, the benzene molecule
adsorbs initially in the butterfly structure but this is observed
to be metastable with respect to a bridging configuration.  They
were able to convert the benzene from one structure to another
using the STM tip and estimated the conversion barrier to be
0.95~eV.  

Gokhale {\it et al} \cite{Gokhale1998a} used thermal desorption
spectroscopy (TPD) and angle-resolved photoelectron spectroscopy
(ARUPS) to investigate the electronic structure and symmetry of
benzene on Si and observed a single dimer structure, supporting
the butterfly configuration.  

Witkowski {\it et al} \cite{Witkowski2003a} used near-edge
x-ray-absoprtion fine-structure (NEXAFS) to look at the structure
of the adsorbed benzene and found the benzene to be symmetric
with respect to the dimer axis, ruling out the tight-bridge 
structure.  These findings were supported by reflectance
anisotropy spectroscopy (RAS) and surface differential reflectivity
spectroscopy (SDRS) data \cite{Witkowski2005a}, which found that
benzene adsorbs on top of a single dimer rather than on the
bridge site between two dimers.  

The high-resolution photoemission study by Kim {\it et al}
\cite{Kim2005a} suggested that the adsorption geometry depends on
the coverage\footnote{In this system the most convenient
  definition for a monolayer (ML) corresponds to one benzene
  molecule per Si dimer.}.  
They found that at low coverages a bridging structure is favoured
but that at high coverages a single dimer structure is more
stable.  This result is consistent with the previous experiments
as the STM measurements were carried out at a low coverage and
the TPD, ARUPS, NEXAFS, RAS and SRDS experiments were carried out
at the saturation coverage of
0.5~ML\cite{Gokhale1998a,Witkowski2005a,Kim2005a}.  

\subsubsection{Adsorption energies}

DFT studies of benzene adsorbed on silicon all agree that the
tight-bridge structure is stable with the butterfly structure
being metastable\cite{Hofer2001a,Lee2005a,Mamatkulov2006a}.  
Lee {\it et al} \cite{Lee2005a} studied the tight-bridge and
butterfly states using VASP and US/norm-conserving (NC)
pseudopotentials.  For a coverage of 0.5~ML the tight-bridge
structure was more stable with an adsorption energy of 1.05~eV.
The adsorption energy of the butterfly structure was 0.82~eV.
They also studied the conversion between the butterfly and
tight-bridge states and found an high energy barrier of 0.87~eV,
which means that both states could coexist.  This supports the
STM findings \cite{Lopinski1998a} but disagrees with other experiments.  
A comparison between the present adsorption energies and those of
previous studies is shown in Table~\ref{tab:benzene-energetics}. 

\begin{table}[ht!]
\begin{center}
\begin{tabular}{|rr|ll|}\hline
\multicolumn{2}{|c|}{Adsorption energy} & & \\
\multicolumn{2}{|c|}{(eV molecule$^{-1}$)} & &  \\
TB   & BF   & Details\footnote{The details are as follows: atomic layers of Si
  in slab, plane-wave cut-off energy, pseudopotential type.  All
  the calculations used either the PW91 or PBE GGAs, which are
  similar.  } & Reference\\ \hline 
1.21 & 0.99 & 9, 400~eV, US    & Present\\
1.26 & 1.01 & 9, 600~eV, PAW   & Present\\
1.25 & 1.00 & 9, 400~eV, PAW   & Present\\
1.21 & 1.02 & 5, 400~eV, PAW   & Present\\
 \hline
0.98 & 0.88 & 6, 350~eV, US    &\cite{Mamatkulov2006a} \\
1.05 & 0.82 & 5, 340~eV, US/NC &\cite{Lee2005a} \\
1.18 & -    & 8, 300~eV, US    &\cite{Hofer2001a}\\
\hline
\end{tabular}
\caption{\label{tab:benzene-energetics} Adsorption energies of
  benzene in the tight-bridge (TB) and butterfly (BF) geometries
  for a coverage of 0.5 ML.  }
\end{center}
\end{table}
In the calculations shown in Table~\ref{tab:benzene-energetics} the
adsorption energies cover a rather large range, which could be
due to different pseudopotentials or supercell size.  All the
results used the PW91 or
PBE\cite{Perdew1996a,Perdew1997a,Perdew1998a} GGA's, which should 
give similar results.  
The current calculations are the most accurate calculations to
date and we have tested the effect of using different slab
thicknesses and pseudopotentials.  Although these make some small
changes to the adsorption energies they do not explain the large
variation seen in the literature and hence we must attribute the
differences to other convergence parameters.  

\subsubsection{Coverage dependence}

Molecular coverage can also have a significant effect on
adsorption as demonstrated by Kim {\it et al} \cite{Kim2005a},
who observed that the structure of the adsorbed benzene is
coverage dependent, with the butterfly structure stable at high
coverages.  Results for a range of coverages are presented in
Tables~\ref{tab:coverage-tb} and \ref{tab:coverage-bf}.  
For the 0.25~ML coverages two supercell orientations are possible.
Supercells (a) and (b) have primitive lattice vectors 
(220) ($\bar{1}10$) (006) and (110) ($\bar{2}$20) (006), respectively, with the
dimer rows along the [$\bar{1}$10] direction.  Supercell (b) has
four dimers along the dimer row.  
\begin{table}[ht!]
\begin{center}
\begin{tabular}{|rr|rrrrr|}\hline
\multicolumn{2}{|c|}{Coverage}&\multicolumn{5}{|c|}{Adsorption energy (eV)} \\
\multicolumn{2}{|c|}{(ML)} & \multicolumn{2}{|c}{Present} & Lee~\cite{Lee2005a} & Hofer~\cite{Hofer2001a}& Jung~\cite{Jung2005a} \\
&& 5-layer & 9-layer    & & & \\ \hline
\multicolumn{2}{|c|}{Isolated} & -    & -    & -    & -    & 0.96 \\
0.125 &     & 1.25 & 1.34 & -    & -    & - \\
0.25  & (a) & 1.23 & 1.27 & -    & 
\multirow{2}{*}{1.42\footnote{Supercell not specified}}& - \\
      & (b) & 1.24 & 1.35 & 0.91 &      & - \\
0.5   &     & 1.21 & 1.25 & 1.05 & 1.18 & - \\
\hline
\end{tabular}
\caption{\label{tab:coverage-tb} Variation of adsorption energy
  with coverage for the tight-bridge structure.  }
\end{center}
\end{table}
\begin{table}[ht!]
\begin{center}
\begin{tabular}{|rr|rrrrr|}\hline
\multicolumn{2}{|c|}{Coverage} & \multicolumn{5}{|c|}{Adsorption energy (eV)} \\
\multicolumn{2}{|c|}{(ML)} & \multicolumn{2}{|c}{Present} & Lee~\cite{Lee2005a} & Hofer~\cite{Hofer2001a}&Jung~\cite{Jung2005a} \\ 
&& 5-layer & 9-layer & & & \\ \hline
\multicolumn{2}{|c|}{Isolated} & -        & -    & -    & -    & 1.04 \\
0.125&     & 1.07     & 1.06 & -    & -    & - \\
0.25 &(a)  & 1.04     & 1.03 & -    &\multirow{2}{*}{1.12\footnote{Supercell not specified}} & - \\
     &(b)  & 1.05     & 1.04 & 0.84 &      & - \\
0.5  &     & 1.02     & 1.00 & 0.82 & -    & - \\
\hline
\end{tabular}
\caption{\label{tab:coverage-bf} Variation of adsorption
  energy with coverage for the butterfly structure.  }
\end{center}
\end{table}

The butterfly structure was found to be unstable at a coverage of
1~ML, which is consistent with the experimental saturation
coverage of around 0.5~ML
\cite{Gokhale1998a,Witkowski2005a,Kim2005a}.  To obtain a 1~ML
coverage for the tight-bridge structure the Si dimers would have
to be shared between the benzene molecules and, thus, for this
case a coverage of 1~ML is unrealistic.  

The present results show a small variation in adsorption
energies for coverages of 0.125-0.5~ML and for all coverages the
tight-bridge structure remains stable.  The adsorption energies
for the butterfly structure do not change significantly when the
slab is increased from 5 to 9 layers.  However for the
tight-bridge structure additional relaxation in the 9-layer
slab results in higher adsorption energies, particularly for the
0.25(b) and 0.125~ML coverages.  

Hofer {\it et al} \cite{Hofer2001a} found that increasing the
coverage from 0.25~ML to 0.5~ML decreased the binding energy of the
tight-bridge structure by 0.24~eV, which was attributed to a
relaxation of the strain within the cell.  This agrees
qualitatively with the present results, which show a smaller
decrease in the tight-bridge adsorption energy of 0.1~eV.  
These results disagree with the DFT results of Lee {\it et al}
\cite{Lee2005a}, who observed an increase of 0.14~eV.  

There are two possible scenarios which could explain the
conflicting experimental and theoretical evidence.  The first
possibility is that the benzene molecule adsorbs initially in the
butterfly structure \cite{Lopinski1998a} (there is no barrier for
this reaction as shown in Figure~\ref{fig:C6H6-barrier}) but this
is metastable with respect to a bridging configuration.  The
conversion barrier was estimated to be 0.95~eV, which is in good
agreement with the DFT calculations of Lee {\it et al}
\cite{Lee2005a}, who found the conversion barrier to be 0.87~eV.  
The barrier is high enough to allow the butterfly configuration
to exist for a relatively long time, which could explain why
other experimental methods have observed the butterfly structure.

The second possibility is that the inclusion of van der Waals
forces will affect the stabilities.  One of the drawbacks of
density functional theory is that it does not include van der
Waals forces, which are due to dynamical correlation.  Even
though benzene binds strongly to Si and the additional binding
due to van der Waals forces will be small compared to the
chemical binding energies, the van der Waals forces may be large
enough to overcome the energy differences between the two
structures and stabilise the butterfly structure.  
The quantum mechanics/molecular mechanics (QM/MM) method used by
Jung {\it et al} \cite{Jung2005a} included van der Waals forces
by using single-point energy calculations with multireference
second-order  perturbation theory (MRMP2).  They found that for
an isolated molecule on a Si cluster the butterfly structure was
more stable than the tight-bridge structure, with adsorption
energies of 1.04 and 0.96~eV, respectively.  However, they used a
small Si cluster, which may not be large enough to represent the
Si surface accurately, so it is not clear whether this result is
valid.  Furthermore, the cluster geometry corresponds to an
isolated molecule i.e. a low coverage system, which according to
Kim {\it et al} is in the regime where the tight-bridge structure
should be stable.  

\subsection{Carbonic acid, H$_2$CO$_3$}

When carbonic acid is placed on the clean Si surface it
dissociates with one O and two H atoms bonding to the Si surface
and a CO$_2$ molecule being left over.  The energy vs. distance
of the H$_2$CO$_3$ approaching the surface was obtained by fixing
the distance of the C atom from the surface and relaxing the
molecule and surface.  This shows that there is no energy barrier
to overcome to enable this reaction to occur.  

\begin{figure}[ht!]
\begin{center}
\includegraphics[height=2.0cm]{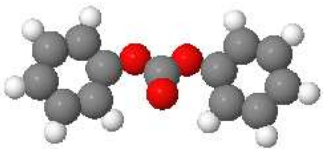}
\caption{\label{fig:C6H5-CO3-C6H5} (Color online) The
C$_6$H$_5$CO$_3$C$_6$H$_5$ molecule represents part of the BPA-PC
monomer containing the carbonate group.  }
\end{center}
\end{figure}

To avoid the problem of dissociation it is necessary to consider
the possible conformations of carbonic acid within the BPA-PC
chain.  The structure of this section of BPA-PC is shown in
Figure~\ref{fig:C6H5-CO3-C6H5}.  

\begin{figure}[ht!]
\begin{center}
\includegraphics[height=3.0cm]{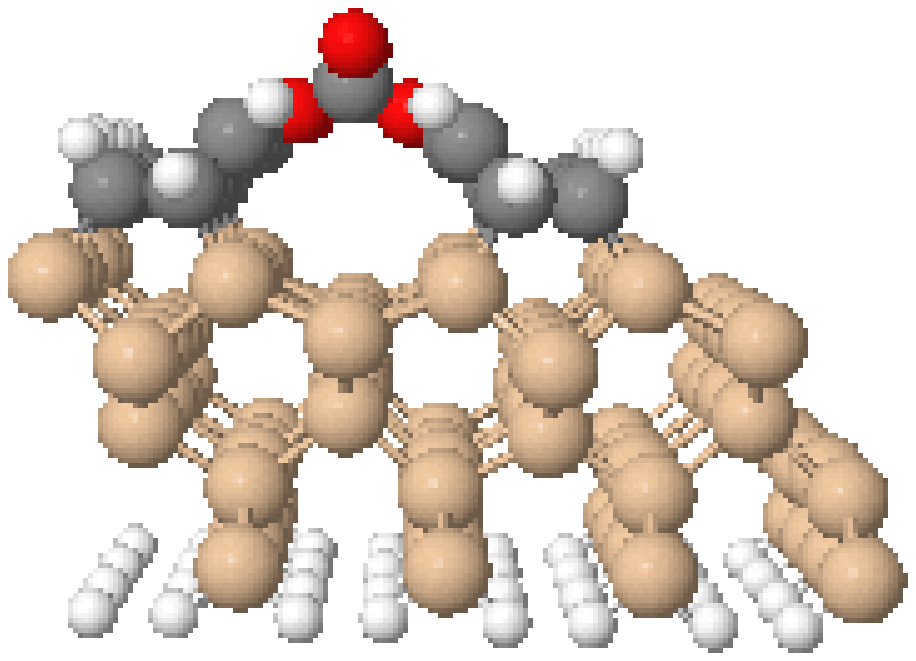}
\includegraphics[height=3.0cm]{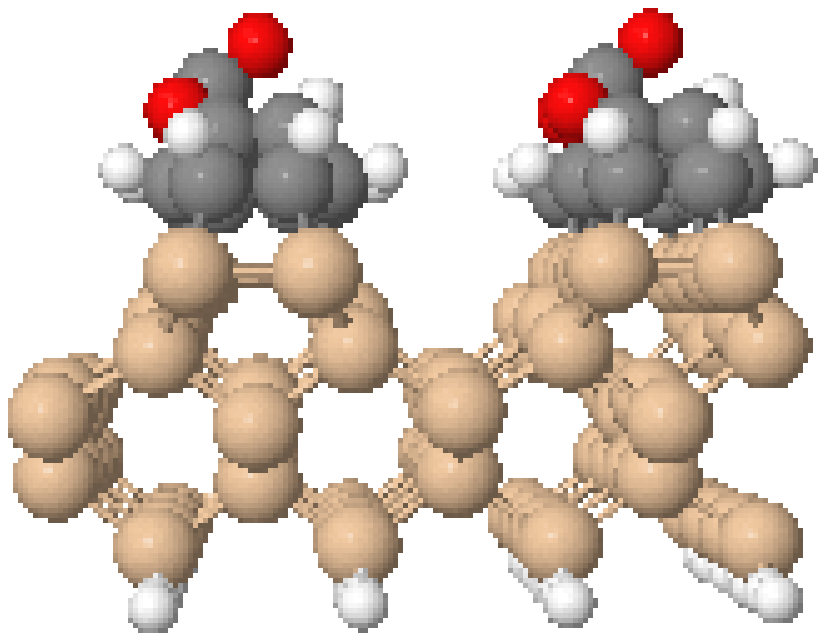}
\caption{\label{fig:C6H5-CO3-C6H5_Si} (Color online) The lowest energy
adsorption structure of C$_6$H$_5$CO$_3$C$_6$H$_5$ adsorbed on
Si.  Both phenol rings have the tight-bridge structure.  LHS:
viewed along [$\bar{1}10$].  RHS: viewed along [110] dimer rows.}
\end{center}
\end{figure}

Based on our knowledge of benzene adsorption on Si the most
stable structure is likely to have the two phenol rings adsorbed
in the tight-bridge structure, as shown in
Figure~\ref{fig:C6H5-CO3-C6H5_Si}.  The molecule is
arranged so that the CO$_3$ group is attached to the top of the
two tight-bridge structures.  Several orientations of the CO$_3$
group within the adsorbed C$_6$H$_5$CO$_3$C$_6$H$_5$ molecule
were calculated and the orientation with the minimum energy has
the C--O double bond pointing upwards and inwards, as shown in
Figure~\ref{fig:C6H5-CO3-C6H5_Si}.  
The adsorption energy of the whole molecule is 1.83~eV.  By
subtracting the adsorption energy of the two tight-bridge
structures from this adsorption energy we estimated the
adsorption energy of the CO$_3$-group to be -0.59~eV.  This means
that the CO$_3$ experiences a repulsive force from the Si
surface.  

Other adsorption configurations for this molecule are unlikely
as the CO$_3$ cannot stretch from a tight-bridge to a butterfly
configuration or from the high point of one tight-bridge to the
low point of an adjacent tight-bridge.  
However, the adsorption barrier for this structure may be high
since due to the geometrical constraints it may be difficult for
both benzenes to initially adsorb in the butterfly structure.  

\subsection{Propane, C$_3$H$_8$}

Similar to the case of carbonic acid, the propane molecule
dissociates on the surface, with the two H atoms bonding to the
surface.  As before, we need to consider a larger segment of
BPA-PC which contains the propane group.  

\begin{figure}[ht!]
\begin{center}
\includegraphics[height=2.2cm,angle=-0]{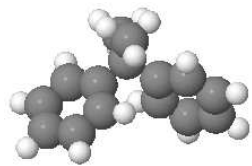}
\caption{\label{fig:C6H5-C3H6-C6H5} (Color online) The isolated
C$_6$H$_5$C$_3$H$_6$C$_6$H$_5$ molecule which is a segment of
BPA-PC containing the propane group.  }
\end{center}
\end{figure}

\begin{figure}[ht!]
\begin{center}
\includegraphics[height=3cm,angle=2]{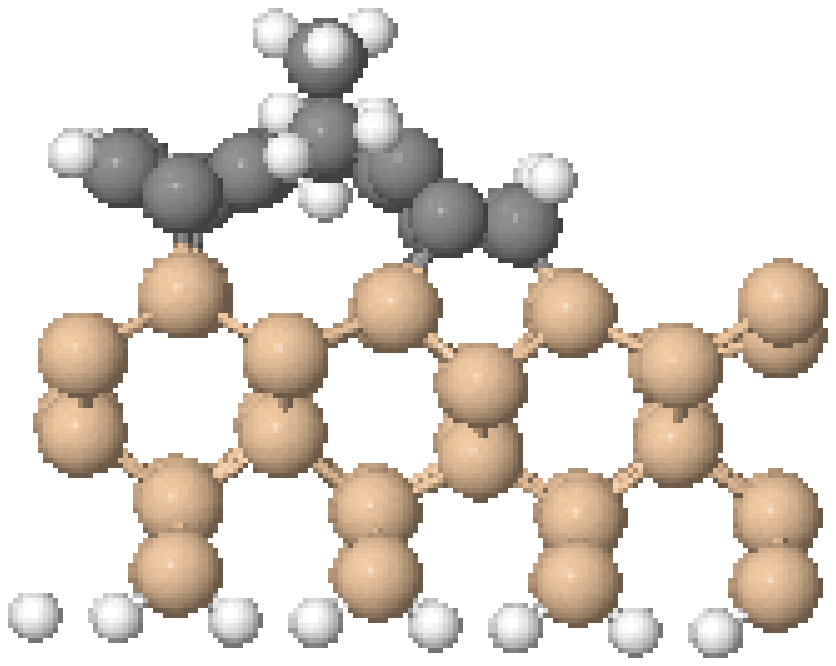}
\includegraphics[height=3cm,angle=1]{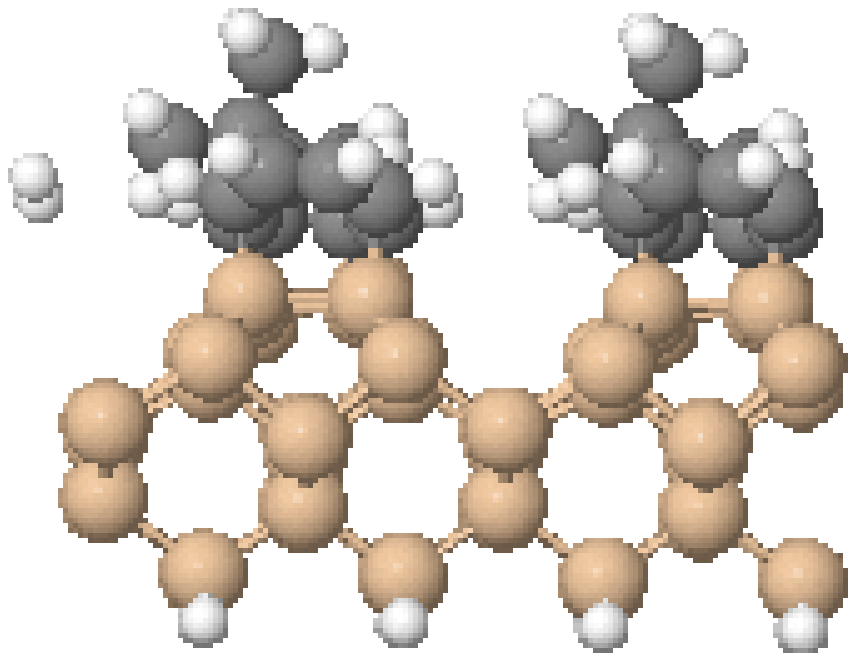}
\caption{\label{fig:C6H5-C3H6-C6H5_Si} (Color online)
C$_6$H$_5$C$_3$H$_6$C$_6$H$_5$ on Si(001)-(2$\times$1) viewed
along [$\bar{1}10$] and [110], respectively.} 
\end{center}
\end{figure}

The isolated molecule is shown in Figure~\ref{fig:C6H5-C3H6-C6H5}
and the molecule adsorbed on the Si surface is shown in
Figure~\ref{fig:C6H5-C3H6-C6H5_Si}.  Geometrically the propane
group cannot bond to two tight-bridge structures but it may bond
to a tight-bridge on one side and the butterfly geometry on the
other.  The lowest energy structure for this configuration is
shown in Figure~\ref{fig:C6H5-C3H6-C6H5_Si} and the adsorption
energy of the whole molecule is 1.59~eV.  The estimated 
adsorption energy of the C$_3$H$_6$-group is -0.64~eV, which
corresponds to a repulsive force from the surface. 

\section{Adsorption on Si(001):H-(2$\times$1)}
\label{sec:H-Si}

\begin{figure}[ht!]
\begin{center}
\includegraphics[height=2.5cm,angle=-90]{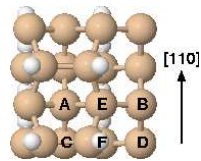}
\caption{\label{fig:sites} (Color online) Adsorption sites on
  Si(001):H-(2$\times$1).  }
\end{center}
\end{figure}

The clean Si surface is very reactive so it is interesting to
know how this compares to the H-passivated surface.  Six
adsorption sites have been considered and are shown in
Figure~\ref{fig:sites}.  Sites A,B,C and D are high symmetry
positions, site F is above a surface H atom and site E is the
midpoint between two surface H atoms.  

\subsection{Benzene, C$_6$H$_6$}

\begin{figure}[ht!]
\begin{center}
\includegraphics[height=3.0cm,angle=0]{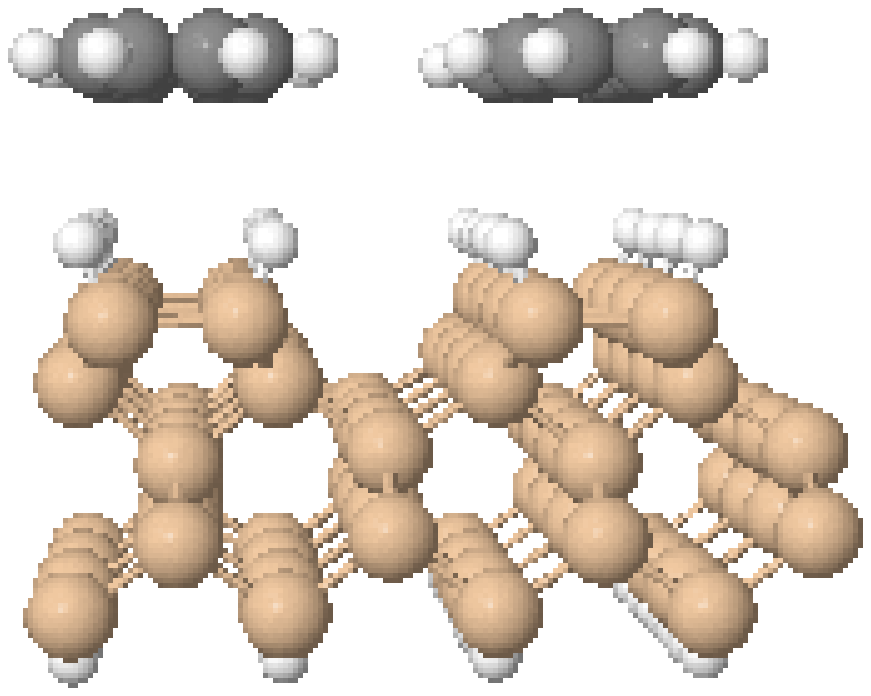}
\includegraphics[height=4.5cm,angle=-0]{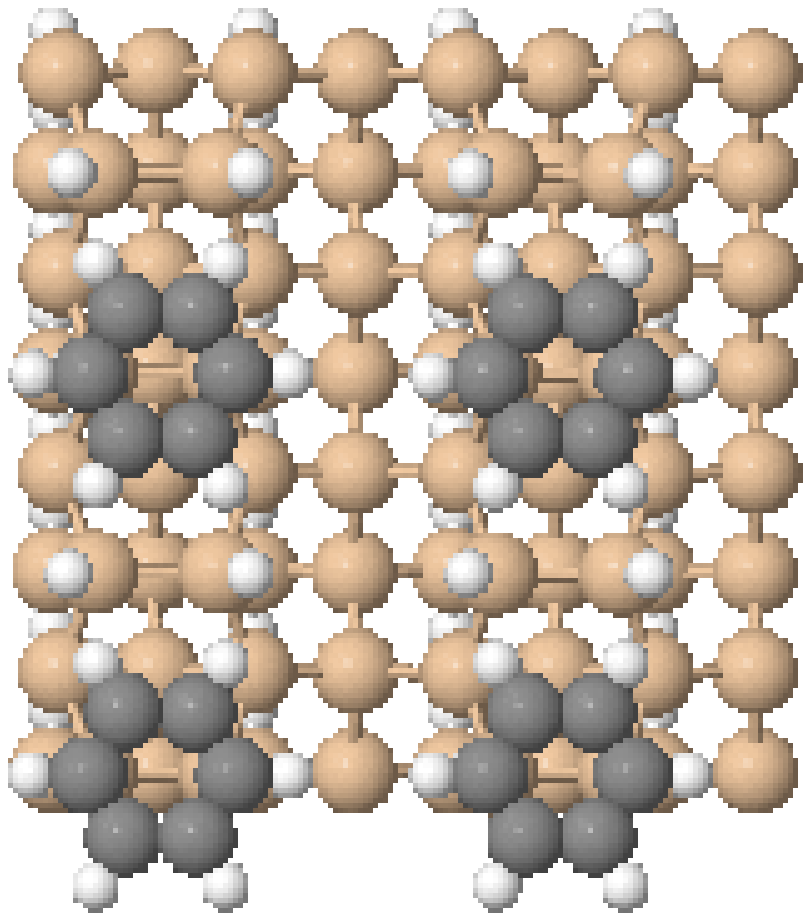}
\caption{\label{fig:adsorption-sites} (Color online) Side and top views
of benzene adsorbed on site C of the Si(001):H-(2$\times$1)
surface and with orientation II.  } 
\end{center}
\end{figure}

There are two orientations for the benzene molecule, denoted I and
II.  Orientation I has two of the C--C bonds in the benzene ring
perpendicular to the Si--Si dimers and orientation II is obtained
by rotating I by 30$^{\rm o}$ around the vertical axis.  As an
example, benzene placed on site C with orientation II, is shown
in Figure~\ref{fig:adsorption-sites}.  

For all sites and orientations the adsorption of benzene is
very weak, with adsorption energies of 0.06-0.07~eV and
equilibrium distances between the benzene and the surface H-atoms
of 3.2-3.4~{\AA}.  The benzene molecule remains flat in contrast
to the large distortion seen in the case on the clean Si surface.

\begin{figure}[ht!]
\begin{center}
\includegraphics[height=5cm]{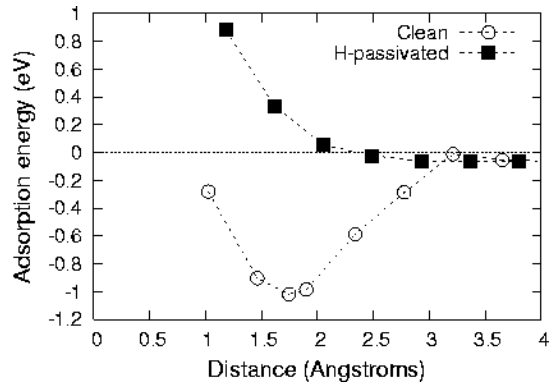}
\caption{\label{fig:C6H6-barrier}
Barriers for adsorption of C$_6$H$_6$ on site CII of
  Si(001)-(2$\times$1) and Si(001):H-(2$\times$1).  }
\end{center}
\end{figure}

The difference between the adsorption behaviour of benzene 
on the clean Si surface and its behaviour on the passivated
surface is further demonstrated by looking at the variation of
the adsorption energy as a function of distance from the
surface, which is shown in Figure~\ref{fig:C6H6-barrier}.  
On the clean Si surface it is known that benzene initially
adsorbs in the butterfly configuration, so for convenience the
benzene was placed in configuration CII on each surface.  For a
particular separation distance the position of one of the C-atoms
was fixed and the rest of the benzene molecule and the surface
were allowed to relax.  
For the clean surface it was convenient to fix either of the C
atoms above the dimer Si atoms and the distance was defined as
the difference between the z-position of the fixed carbon atom
and the average z-position of the two Si dimer atoms in the
minimum energy configuration.  
For the passivated surface the distance was defined as the
distance from the fixed carbon atom to the z-position of the
surface H-atoms at the minimum energy configuration.  The
position of the H atoms is not significantly different for the
various benzene configurations.  
As can be seen from Figure~\ref{fig:C6H6-barrier} there is no
barrier for adsorption into the butterfly configuration on the
clean Si surface.  For the H-passivated surface, the benzene
experiences significant repulsion as it approaches the surface
and thus the benzene sees the surface as a approximately uniform
hard wall.  

The binding energies for the passivated surface are small
and therefore to get the true adsorption energies and equilibrium
distances of these molecules it is necessary to consider van der
Waals interactions.  Based on calculations
of phenol on alumina \cite{Chakarova2006a} and phenol on graphite
\cite{Chakarova2006b} we estimate the van der Waals energy to be
of the order of 0.2-0.5~eV.   

\subsection{Carbonic acid, H$_2$CO$_3$}

\begin{figure}[ht!]
\begin{center}
\includegraphics[height=2.8cm]{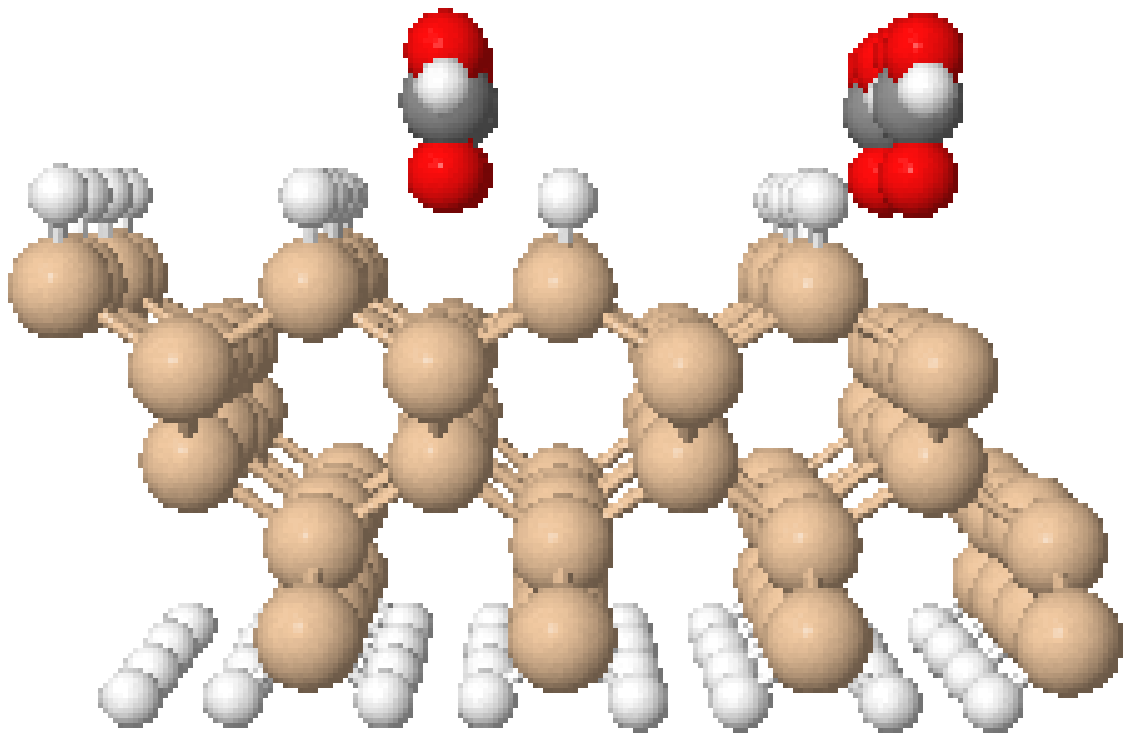}
\includegraphics[height=2.8cm]{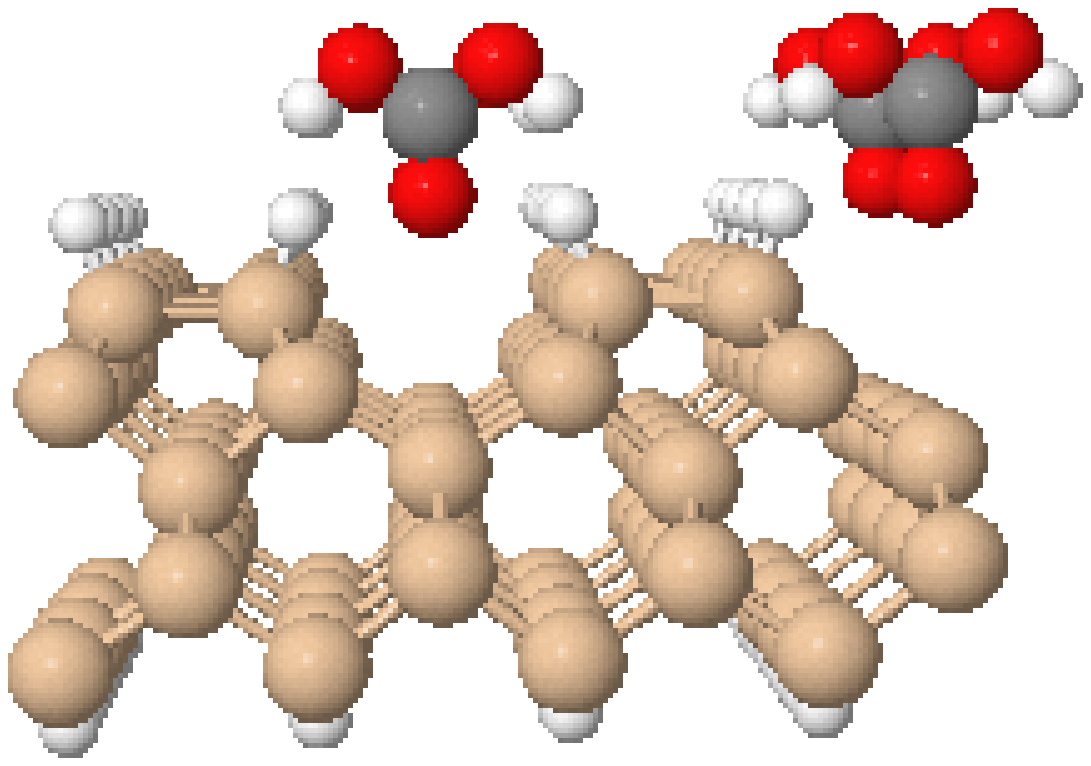}
\caption{\label{fig:h2co3_hsi_b3} (Color online) Carbonic acid molecule
adsorbed on the H-passivated Si surface with configuration
BIII. LHS: viewed along [$\bar{1}10$].  RHS: viewed along [110]
dimer rows. }
\end{center}
\end{figure}

There are four possible orientations for the H$_2$CO$_3$ molecule to
adsorb on the H-passivated Si(001) surface.  These are I) parallel 
to a dimer with the double-bonded O on top, II) perpendicular to
a dimer with the double-bonded O on top, and III) parallel with
the double-bonded O pointing down and IV) perpendicular with the
double-bonded O pointing down.  Configuration BIII is shown in
Figure~\ref{fig:h2co3_hsi_b3}.   

The binding energies for all the sites range from 0.04-0.10~eV
and the distance of the carbon atom in H$_2$CO$_3$ from the
surface hydrogen atoms ranges from 1.6-4.2~{\AA}.  
The maximum binding energy and minimum adsorption distance occurs
for configuration BIII and as can be seen from
Figure~\ref{fig:h2co3_hsi_b3} this is mainly due to topology.

\subsection{Propane, C$_3$H$_8$}

\begin{figure}[ht!]
\begin{center}
\includegraphics[height=3.0cm]{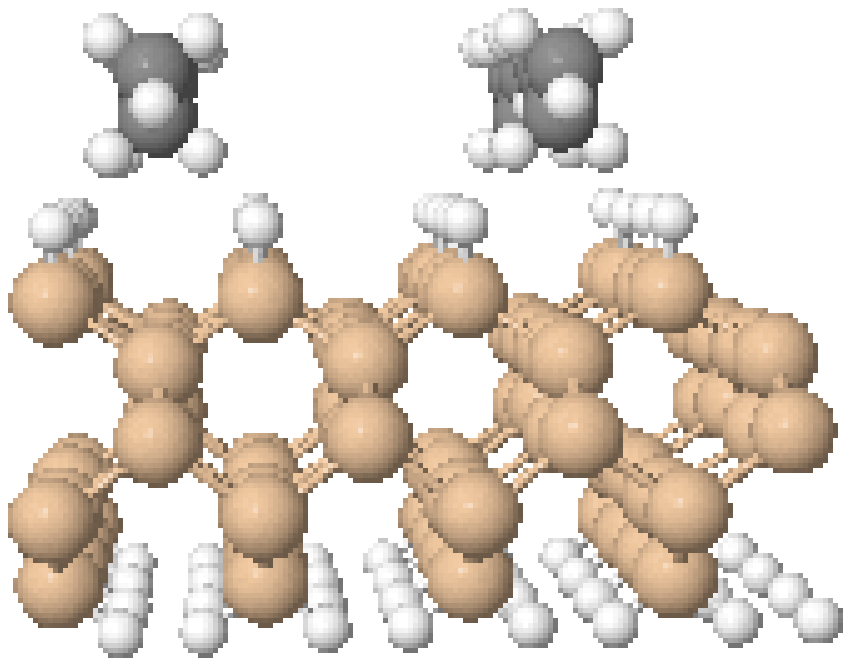}
\includegraphics[height=3.0cm]{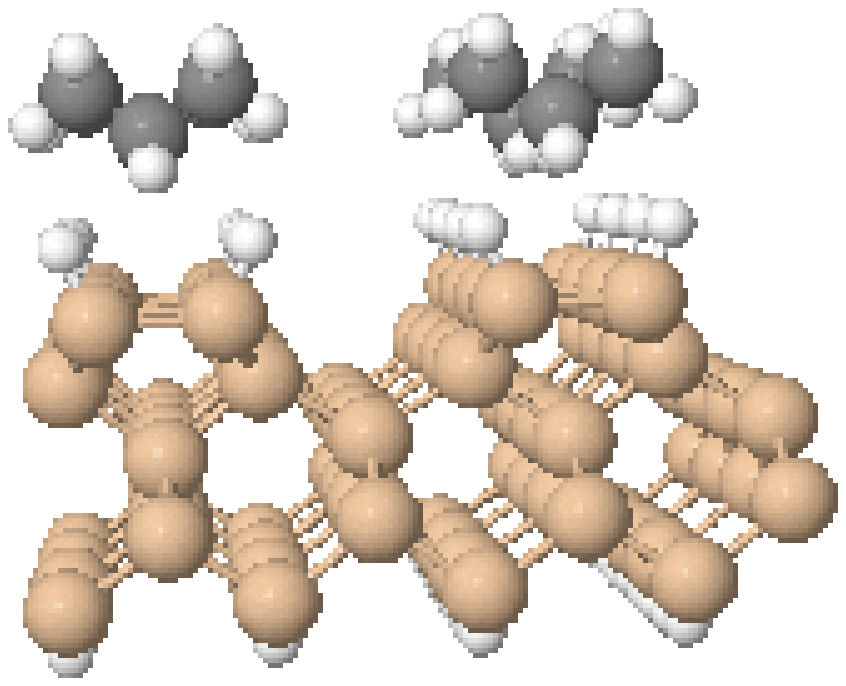}
\caption{\label{fig:c3h8_hsi_c1} (Color online) Propane molecule
adsorbed on the H-passivated Si surface with configuration CI.
LHS: viewed along [$\bar{1}10$].  RHS: viewed along [110] dimer rows.}
\end{center}
\end{figure}

There are four possible orientations in which propane could
adsorb.  In I and III the carbon chain is parallel to the dimers,
whereas in II and IV the carbon chain is perpendicular to the
dimers.  For I and II the carbon-chain makes a ``V-shape'' on
the surface and III and IV the carbon-chain is inverted to make
an upside down ``V''.  Figure~\ref{fig:c3h8_hsi_c1} shows the
propane molecule in site C and orientation I, in which the
``V-shape'' can be seen on the right.  
Similar to benzene and carbonic acid, propane is also very weakly
bound to the surface with adsorption energies ranging from
0.04-0.06~eV.  For orientations I and II the vertical distance
from the central C in propane to the surface H-atoms is
2.5-3.4~{\AA} and for orientations III and IV it is
3.8-4.2~{\AA}.  

\section{Summary and Discussion}
\label{sec:summary}

The adsorption behaviour of benzene, carbonic acid and propane
on the Si(001)-(2$\times$1) and Si(001):H-(2$\times$1) surfaces
was calculated.  
For the Si(001)-(2$\times$1) surface the stable adsorption
structure of benzene was found to be the tight-bridge structure
with an adsorption energy of 1.26~eV.  The butterfly structure is
metastable with an adsorption energy of 1.01~eV.  The carbonic
acid and propane molecules were found to spontaneously dissociate
on the surface.  

To avoid the issue of dissociation, the adsorption behaviour of
larger segments of the BPA-PC chain was studied.  These
segments correspond to carbonic acid bonded to two benzene rings
and propane bonded to two benzene rings.  Trial structures of
these large molecules were calculated and the results were used
to estimate adsorption energies of the CO$_3$ and C$_3$H$_6$
groups.  It was found that these groups are repelled from the
Si surface.  

Combined DFT/coarse-grained studies of BPA-PC on Ni(111) have
already been published \cite{Dellesite2002a,Abrams2003a}.  
The first-principles calculations showed that benzene binds
strongly to the Ni surface but that carbonic acid and propane
experience significant repulsion.  
The behaviour of these molecules on Si is very different
to their behaviour on Ni(111).  The adsorption energy of benzene
on Ni is 1.05~eV \cite{Dellesite2002a}, which is comparable to the
adsorption energy of benzene on Si.  The geometries, however, are
very different, and on the Ni surface the benzene remains
relatively undistorted at a centre-of-mass distance of 2~{\AA}. 
In contrast, propane and carbonic acid see the Ni
surface as a uniform hard wall and they experience significant
repulsion for distances shorter than 3.2~{\AA}.  
By incorporating this data into a coarse-grained model, they
showed that the phenol chain ends bind to the Ni surface and that
the rest of the polymer is repelled away from the surface.  

An important point to note is that the trial adsorption
structures for the BPA-PC segments, described above, may not
be geometrically compatible with the larger BPA-PC chain.
The polymer chain must be continued from the low point of
the tight-bridge structure and if the repulsion of the
C$_3$H$_6$ group is strong enough this benzene may
desorb from the surface.  In this case the ``internal'' benzene
groups may not approach the surface and only the benzene chain
ends may adsorb.  If this is the case, then despite the
differences in molecular adsorption behaviour the  behaviour of a
BPA-PC chain on Si may turn out to be similar to that of BPA-PC
on Ni.  

Passivation of the Si surface with H-atoms has a dramatic effect on
the surface properties.  The passivated surface is very inert and
the binding energy of all the molecules is extremely weak.  Due to
the weak binding of this system the van der Waals interaction
becomes dominant and to get a true picture of the binding in this
case it would be necessary to include the van der Waals energy.  

Another physical effect that is missing in the DFT formalism is
the image charge potential.  This has been addressed in a recent
study of adsorbed molecules on graphite \cite{Neaton2006a}.
The image charge affects, in particular, the excited-state energy
levels but not the adsorption energies.  

This work provides the first step in multiscale simulations of the
adhesion of polymers to surfaces.  
The first-principles calculations presented in this paper can now
be used to build potentials for coarse-grained simulations.

\begin{acknowledgments}
The authors would like to thank Janne Blomqvist, Petri Salo,
Jyrki Vuorinen and
Luigi Delle Site for useful discussions.  Support was provided by
the Finnish Funding Agency for Technology and Innovation (TEKES).
Computational resources were provided by the Computer Science
Center (CSC).  
\end{acknowledgments}

\bibliography{hypris}

\begin{thebibliography}{36}
\expandafter\ifx\csname natexlab\endcsname\relax\def\natexlab#1{#1}\fi
\expandafter\ifx\csname bibnamefont\endcsname\relax
  \def\bibnamefont#1{#1}\fi
\expandafter\ifx\csname bibfnamefont\endcsname\relax
  \def\bibfnamefont#1{#1}\fi
\expandafter\ifx\csname citenamefont\endcsname\relax
  \def\citenamefont#1{#1}\fi
\expandafter\ifx\csname url\endcsname\relax
  \def\url#1{\texttt{#1}}\fi
\expandafter\ifx\csname urlprefix\endcsname\relax\def\urlprefix{URL }\fi
\providecommand{\bibinfo}[2]{#2}
\providecommand{\eprint}[2][]{\url{#2}}

\bibitem[{\citenamefont{Lopinski et~al.}(2000)\citenamefont{Lopinski, Wayner,
  and Wolkow}}]{Lopinski2000a}
\bibinfo{author}{\bibfnamefont{G.}~\bibnamefont{Lopinski}},
  \bibinfo{author}{\bibfnamefont{D.}~\bibnamefont{Wayner}}, \bibnamefont{and}
  \bibinfo{author}{\bibfnamefont{R.}~\bibnamefont{Wolkow}},
  \bibinfo{journal}{Nature} \textbf{\bibinfo{volume}{406}}, \bibinfo{pages}{48}
  (\bibinfo{year}{2000}).

\bibitem[{\citenamefont{Besley and Blundy}(2006)}]{Besley2006a}
\bibinfo{author}{\bibfnamefont{N.~A.} \bibnamefont{Besley}} \bibnamefont{and}
  \bibinfo{author}{\bibfnamefont{A.~J.} \bibnamefont{Blundy}},
  \bibinfo{journal}{J. Phys. Chem. B} \textbf{\bibinfo{volume}{110}},
  \bibinfo{pages}{1701} (\bibinfo{year}{2006}).

\bibitem[{\citenamefont{Quek et~al.}(2006)\citenamefont{Quek, Neaton,
  Hybertsen, Kaxiras, and Louie}}]{Quek2006a}
\bibinfo{author}{\bibfnamefont{S.~Y.} \bibnamefont{Quek}},
  \bibinfo{author}{\bibfnamefont{J.~B.} \bibnamefont{Neaton}},
  \bibinfo{author}{\bibfnamefont{M.~S.} \bibnamefont{Hybertsen}},
  \bibinfo{author}{\bibfnamefont{E.}~\bibnamefont{Kaxiras}}, \bibnamefont{and}
  \bibinfo{author}{\bibfnamefont{S.~G.} \bibnamefont{Louie}},
  \bibinfo{journal}{Phys. Stat. Sol. (b)} \textbf{\bibinfo{volume}{243}},
  \bibinfo{pages}{2048} (\bibinfo{year}{2006}).

\bibitem[{\citenamefont{Montanari
  et~al.}(1998{\natexlab{a}})\citenamefont{Montanari, Ballone, and
  Jones}}]{Montanari1998a}
\bibinfo{author}{\bibfnamefont{B.}~\bibnamefont{Montanari}},
  \bibinfo{author}{\bibfnamefont{P.}~\bibnamefont{Ballone}}, \bibnamefont{and}
  \bibinfo{author}{\bibfnamefont{R.}~\bibnamefont{Jones}}, \bibinfo{journal}{J.
  Chem. Phys.} \textbf{\bibinfo{volume}{108}}, \bibinfo{pages}{6947}
  (\bibinfo{year}{1998}{\natexlab{a}}).

\bibitem[{\citenamefont{Montanari
  et~al.}(1998{\natexlab{b}})\citenamefont{Montanari, Ballone, and
  Jones}}]{Montanari1998b}
\bibinfo{author}{\bibfnamefont{B.}~\bibnamefont{Montanari}},
  \bibinfo{author}{\bibfnamefont{P.}~\bibnamefont{Ballone}}, \bibnamefont{and}
  \bibinfo{author}{\bibfnamefont{R.}~\bibnamefont{Jones}},
  \bibinfo{journal}{Macromolecules} \textbf{\bibinfo{volume}{31}},
  \bibinfo{pages}{7784} (\bibinfo{year}{1998}{\natexlab{b}}).

\bibitem[{\citenamefont{Montanari et~al.}(1999)\citenamefont{Montanari,
  Ballone, and Jones}}]{Montanari1999a}
\bibinfo{author}{\bibfnamefont{B.}~\bibnamefont{Montanari}},
  \bibinfo{author}{\bibfnamefont{P.}~\bibnamefont{Ballone}}, \bibnamefont{and}
  \bibinfo{author}{\bibfnamefont{R.}~\bibnamefont{Jones}},
  \bibinfo{journal}{Macromolecules} \textbf{\bibinfo{volume}{32}},
  \bibinfo{pages}{3396} (\bibinfo{year}{1999}).

\bibitem[{\citenamefont{Abrams and Kremer}(2003)}]{Abrams2003b}
\bibinfo{author}{\bibfnamefont{C.~F.} \bibnamefont{Abrams}} \bibnamefont{and}
  \bibinfo{author}{\bibfnamefont{K.}~\bibnamefont{Kremer}},
  \bibinfo{journal}{Macromolecules} \textbf{\bibinfo{volume}{36}},
  \bibinfo{pages}{260} (\bibinfo{year}{2003}).

\bibitem[{\citenamefont{Dion et~al.}(2004)\citenamefont{Dion, Rydberg,
  Schr\"{o}der, Langreth, and Lundqvist}}]{Dion2004a}
\bibinfo{author}{\bibfnamefont{M.}~\bibnamefont{Dion}},
  \bibinfo{author}{\bibfnamefont{H.}~\bibnamefont{Rydberg}},
  \bibinfo{author}{\bibfnamefont{E.}~\bibnamefont{Schr\"{o}der}},
  \bibinfo{author}{\bibfnamefont{D.}~\bibnamefont{Langreth}}, \bibnamefont{and}
  \bibinfo{author}{\bibfnamefont{B.}~\bibnamefont{Lundqvist}},
  \bibinfo{journal}{Phys. Rev. Lett.} \textbf{\bibinfo{volume}{92}},
  \bibinfo{pages}{246401} (\bibinfo{year}{2004}).

\bibitem[{\citenamefont{Dion et~al.}(2005)\citenamefont{Dion, Rydberg,
  Schr\"{o}der, Langreth, and Lundqvist}}]{Dion2005a}
\bibinfo{author}{\bibfnamefont{M.}~\bibnamefont{Dion}},
  \bibinfo{author}{\bibfnamefont{H.}~\bibnamefont{Rydberg}},
  \bibinfo{author}{\bibfnamefont{E.}~\bibnamefont{Schr\"{o}der}},
  \bibinfo{author}{\bibfnamefont{D.}~\bibnamefont{Langreth}}, \bibnamefont{and}
  \bibinfo{author}{\bibfnamefont{B.}~\bibnamefont{Lundqvist}},
  \bibinfo{journal}{Phys. Rev. Lett.} \textbf{\bibinfo{volume}{95}},
  \bibinfo{pages}{109902} (\bibinfo{year}{2005}).

\bibitem[{\citenamefont{Puzder et~al.}(2006)\citenamefont{Puzder, Dion, and
  Langreth}}]{Puzder2006a}
\bibinfo{author}{\bibfnamefont{A.}~\bibnamefont{Puzder}},
  \bibinfo{author}{\bibfnamefont{M.}~\bibnamefont{Dion}}, \bibnamefont{and}
  \bibinfo{author}{\bibfnamefont{D.~C.} \bibnamefont{Langreth}},
  \bibinfo{journal}{J. Chem. Phys.} \textbf{\bibinfo{volume}{124}},
  \bibinfo{pages}{164105} (\bibinfo{year}{2006}).

\bibitem[{\citenamefont{Thonhauser et~al.}(2006)\citenamefont{Thonhauser,
  Puzder, and Langreth}}]{Thonhauser2006a}
\bibinfo{author}{\bibfnamefont{T.}~\bibnamefont{Thonhauser}},
  \bibinfo{author}{\bibfnamefont{A.}~\bibnamefont{Puzder}}, \bibnamefont{and}
  \bibinfo{author}{\bibfnamefont{D.~C.} \bibnamefont{Langreth}},
  \bibinfo{journal}{J. Chem. Phys.} \textbf{\bibinfo{volume}{124}},
  \bibinfo{pages}{164106} (\bibinfo{year}{2006}).

\bibitem[{\citenamefont{Chakarova-K\"{a}ck
  et~al.}(2006{\natexlab{a}})\citenamefont{Chakarova-K\"{a}ck, Schr\"{o}der,
  Lundqvist, and Langreth}}]{Chakarova2006a}
\bibinfo{author}{\bibfnamefont{S.~D.} \bibnamefont{Chakarova-K\"{a}ck}},
  \bibinfo{author}{\bibfnamefont{E.}~\bibnamefont{Schr\"{o}der}},
  \bibinfo{author}{\bibfnamefont{B.~I.} \bibnamefont{Lundqvist}},
  \bibnamefont{and} \bibinfo{author}{\bibfnamefont{D.~C.}
  \bibnamefont{Langreth}}, \bibinfo{journal}{Phys. Rev. Lett.}
  \textbf{\bibinfo{volume}{96}}, \bibinfo{pages}{146107}
  (\bibinfo{year}{2006}{\natexlab{a}}).

\bibitem[{\citenamefont{Chakarova-K\"{a}ck
  et~al.}(2006{\natexlab{b}})\citenamefont{Chakarova-K\"{a}ck, Borck,
  Schr\"{o}der, and Lundqvist}}]{Chakarova2006b}
\bibinfo{author}{\bibfnamefont{S.~D.} \bibnamefont{Chakarova-K\"{a}ck}},
  \bibinfo{author}{\bibfnamefont{O.}~\bibnamefont{Borck}},
  \bibinfo{author}{\bibfnamefont{E.}~\bibnamefont{Schr\"{o}der}},
  \bibnamefont{and} \bibinfo{author}{\bibfnamefont{B.~I.}
  \bibnamefont{Lundqvist}}, \bibinfo{journal}{Phys. Rev. B}
  \textbf{\bibinfo{volume}{74}}, \bibinfo{pages}{155402}
  (\bibinfo{year}{2006}{\natexlab{b}}).

\bibitem[{\citenamefont{Kresse and
  Furthm\"{u}ller}(1996{\natexlab{a}})}]{Kresse1996a}
\bibinfo{author}{\bibfnamefont{G.}~\bibnamefont{Kresse}} \bibnamefont{and}
  \bibinfo{author}{\bibfnamefont{J.}~\bibnamefont{Furthm\"{u}ller}},
  \bibinfo{journal}{Comput. Mat. Sci.} \textbf{\bibinfo{volume}{6}},
  \bibinfo{pages}{15} (\bibinfo{year}{1996}{\natexlab{a}}).

\bibitem[{\citenamefont{Kresse and
  Furthm\"{u}ller}(1996{\natexlab{b}})}]{Kresse1996b}
\bibinfo{author}{\bibfnamefont{G.}~\bibnamefont{Kresse}} \bibnamefont{and}
  \bibinfo{author}{\bibfnamefont{J.}~\bibnamefont{Furthm\"{u}ller}},
  \bibinfo{journal}{Phys. Rev. B} \textbf{\bibinfo{volume}{54}},
  \bibinfo{pages}{11169} (\bibinfo{year}{1996}{\natexlab{b}}).

\bibitem[{\citenamefont{Perdew}(1991)}]{Perdew1991a}
\bibinfo{author}{\bibfnamefont{J.}~\bibnamefont{Perdew}},
  \emph{\bibinfo{title}{Electronic Structure of Solids '91}}
  (\bibinfo{publisher}{Academie Verlag}, \bibinfo{year}{1991}),
  p.~\bibinfo{pages}{11}.

\bibitem[{\citenamefont{Perdew et~al.}(1996{\natexlab{a}})\citenamefont{Perdew,
  Burke, and Wang}}]{Perdew1996b}
\bibinfo{author}{\bibfnamefont{J.}~\bibnamefont{Perdew}},
  \bibinfo{author}{\bibfnamefont{K.}~\bibnamefont{Burke}}, \bibnamefont{and}
  \bibinfo{author}{\bibfnamefont{Y.}~\bibnamefont{Wang}},
  \bibinfo{journal}{Phys. Rev. B} \textbf{\bibinfo{volume}{54}},
  \bibinfo{pages}{16533} (\bibinfo{year}{1996}{\natexlab{a}}).

\bibitem[{\citenamefont{Bl\"{o}chl}(1994)}]{Blochl1994a}
\bibinfo{author}{\bibfnamefont{P.}~\bibnamefont{Bl\"{o}chl}},
  \bibinfo{journal}{Phys. Rev. B} \textbf{\bibinfo{volume}{50}},
  \bibinfo{pages}{17953} (\bibinfo{year}{1994}).

\bibitem[{\citenamefont{Zanella et~al.}(2006)\citenamefont{Zanella, Fazzio, and
  da~Silva}}]{Zanella2006a}
\bibinfo{author}{\bibfnamefont{I.}~\bibnamefont{Zanella}},
  \bibinfo{author}{\bibfnamefont{A.}~\bibnamefont{Fazzio}}, \bibnamefont{and}
  \bibinfo{author}{\bibfnamefont{A.~J.~R.} \bibnamefont{da~Silva}},
  \bibinfo{journal}{J. Phys. Chem. B} \textbf{\bibinfo{volume}{110}},
  \bibinfo{pages}{10849} (\bibinfo{year}{2006}).

\bibitem[{\citenamefont{Hofer et~al.}(2001)\citenamefont{Hofer, Fisher,
  Lopinski, and Wolkow}}]{Hofer2001a}
\bibinfo{author}{\bibfnamefont{W.}~\bibnamefont{Hofer}},
  \bibinfo{author}{\bibfnamefont{A.}~\bibnamefont{Fisher}},
  \bibinfo{author}{\bibfnamefont{G.}~\bibnamefont{Lopinski}}, \bibnamefont{and}
  \bibinfo{author}{\bibfnamefont{R.}~\bibnamefont{Wolkow}},
  \bibinfo{journal}{Phys. Rev. B} \textbf{\bibinfo{volume}{63}},
  \bibinfo{pages}{85314} (\bibinfo{year}{2001}).

\bibitem[{\citenamefont{Jung and Gordon}(2005)}]{Jung2005a}
\bibinfo{author}{\bibfnamefont{Y.}~\bibnamefont{Jung}} \bibnamefont{and}
  \bibinfo{author}{\bibfnamefont{M.~S.} \bibnamefont{Gordon}},
  \bibinfo{journal}{J. Am. Chem. Soc.} \textbf{\bibinfo{volume}{127}},
  \bibinfo{pages}{3131} (\bibinfo{year}{2005}).

\bibitem[{\citenamefont{Wolkow et~al.}(1998)\citenamefont{Wolkow, Lopinski, and
  Moffatt}}]{Wolkow1998a}
\bibinfo{author}{\bibfnamefont{R.}~\bibnamefont{Wolkow}},
  \bibinfo{author}{\bibfnamefont{G.}~\bibnamefont{Lopinski}}, \bibnamefont{and}
  \bibinfo{author}{\bibfnamefont{D.}~\bibnamefont{Moffatt}},
  \bibinfo{journal}{Surf. Sci.} \textbf{\bibinfo{volume}{416}},
  \bibinfo{pages}{L1107} (\bibinfo{year}{1998}).

\bibitem[{\citenamefont{Birkenheuer et~al.}(1998)\citenamefont{Birkenheuer,
  Gutdeutsch, and R\"{o}sch}}]{Birkenheuer1998a}
\bibinfo{author}{\bibfnamefont{U.}~\bibnamefont{Birkenheuer}},
  \bibinfo{author}{\bibfnamefont{U.}~\bibnamefont{Gutdeutsch}},
  \bibnamefont{and}
  \bibinfo{author}{\bibfnamefont{N.}~\bibnamefont{R\"{o}sch}},
  \bibinfo{journal}{Surf. Sci.} \textbf{\bibinfo{volume}{409}},
  \bibinfo{pages}{213} (\bibinfo{year}{1998}).

\bibitem[{\citenamefont{Lopinski et~al.}(1998)\citenamefont{Lopinski, Moffatt,
  and Wolkow}}]{Lopinski1998a}
\bibinfo{author}{\bibfnamefont{G.}~\bibnamefont{Lopinski}},
  \bibinfo{author}{\bibfnamefont{D.}~\bibnamefont{Moffatt}}, \bibnamefont{and}
  \bibinfo{author}{\bibfnamefont{R.}~\bibnamefont{Wolkow}},
  \bibinfo{journal}{Chem. Phys. Lett.} \textbf{\bibinfo{volume}{282}},
  \bibinfo{pages}{305} (\bibinfo{year}{1998}).

\bibitem[{\citenamefont{Gokhale et~al.}(1998)\citenamefont{Gokhale,
  Trischberger, Menzel, Widdra, Dr\"{o}ge, Steinr\"{u}ck, Birkenheuer,
  Gutdeutsch, and R\"{o}sch}}]{Gokhale1998a}
\bibinfo{author}{\bibfnamefont{S.}~\bibnamefont{Gokhale}},
  \bibinfo{author}{\bibfnamefont{P.}~\bibnamefont{Trischberger}},
  \bibinfo{author}{\bibfnamefont{D.}~\bibnamefont{Menzel}},
  \bibinfo{author}{\bibfnamefont{W.}~\bibnamefont{Widdra}},
  \bibinfo{author}{\bibfnamefont{H.}~\bibnamefont{Dr\"{o}ge}},
  \bibinfo{author}{\bibfnamefont{H.-P.} \bibnamefont{Steinr\"{u}ck}},
  \bibinfo{author}{\bibfnamefont{U.}~\bibnamefont{Birkenheuer}},
  \bibinfo{author}{\bibfnamefont{U.}~\bibnamefont{Gutdeutsch}},
  \bibnamefont{and}
  \bibinfo{author}{\bibfnamefont{N.}~\bibnamefont{R\"{o}sch}},
  \bibinfo{journal}{J. Chem. Phys.} \textbf{\bibinfo{volume}{108}},
  \bibinfo{pages}{5554} (\bibinfo{year}{1998}).

\bibitem[{\citenamefont{Witkowski et~al.}(2003)\citenamefont{Witkowski,
  Hennies, Pietzsch, Mattsson, F\"{o}hlisch, Wurth, Nagasono, and
  Piancastelli}}]{Witkowski2003a}
\bibinfo{author}{\bibfnamefont{N.}~\bibnamefont{Witkowski}},
  \bibinfo{author}{\bibfnamefont{F.}~\bibnamefont{Hennies}},
  \bibinfo{author}{\bibfnamefont{A.}~\bibnamefont{Pietzsch}},
  \bibinfo{author}{\bibfnamefont{S.}~\bibnamefont{Mattsson}},
  \bibinfo{author}{\bibfnamefont{A.}~\bibnamefont{F\"{o}hlisch}},
  \bibinfo{author}{\bibfnamefont{W.}~\bibnamefont{Wurth}},
  \bibinfo{author}{\bibfnamefont{M.}~\bibnamefont{Nagasono}}, \bibnamefont{and}
  \bibinfo{author}{\bibfnamefont{M.}~\bibnamefont{Piancastelli}},
  \bibinfo{journal}{Phys. Rev. B} \textbf{\bibinfo{volume}{68}},
  \bibinfo{pages}{115408} (\bibinfo{year}{2003}).

\bibitem[{\citenamefont{Witkowski et~al.}(2005)\citenamefont{Witkowski,
  Pluchery, and Borensztein}}]{Witkowski2005a}
\bibinfo{author}{\bibfnamefont{N.}~\bibnamefont{Witkowski}},
  \bibinfo{author}{\bibfnamefont{O.}~\bibnamefont{Pluchery}}, \bibnamefont{and}
  \bibinfo{author}{\bibfnamefont{Y.}~\bibnamefont{Borensztein}},
  \bibinfo{journal}{Phys. Rev. B} \textbf{\bibinfo{volume}{72}},
  \bibinfo{pages}{75354} (\bibinfo{year}{2005}).

\bibitem[{\citenamefont{Kim et~al.}(2005)\citenamefont{Kim, Lee, and
  Yeom}}]{Kim2005a}
\bibinfo{author}{\bibfnamefont{Y.}~\bibnamefont{Kim}},
  \bibinfo{author}{\bibfnamefont{M.}~\bibnamefont{Lee}}, \bibnamefont{and}
  \bibinfo{author}{\bibfnamefont{H.}~\bibnamefont{Yeom}},
  \bibinfo{journal}{Phys. Rev. B} \textbf{\bibinfo{volume}{71}},
  \bibinfo{pages}{115311} (\bibinfo{year}{2005}).

\bibitem[{\citenamefont{Lee and Cho}(2005)}]{Lee2005a}
\bibinfo{author}{\bibfnamefont{J.-Y.} \bibnamefont{Lee}} \bibnamefont{and}
  \bibinfo{author}{\bibfnamefont{J.-H.} \bibnamefont{Cho}},
  \bibinfo{journal}{Phys. Rev. B} \textbf{\bibinfo{volume}{72}},
  \bibinfo{pages}{235317} (\bibinfo{year}{2005}).

\bibitem[{\citenamefont{Mamatkulov et~al.}(2006)\citenamefont{Mamatkulov,
  Stauffer, Minot, and Sonnet}}]{Mamatkulov2006a}
\bibinfo{author}{\bibfnamefont{M.}~\bibnamefont{Mamatkulov}},
  \bibinfo{author}{\bibfnamefont{L.}~\bibnamefont{Stauffer}},
  \bibinfo{author}{\bibfnamefont{C.}~\bibnamefont{Minot}}, \bibnamefont{and}
  \bibinfo{author}{\bibfnamefont{P.}~\bibnamefont{Sonnet}},
  \bibinfo{journal}{Phys. Rev. B} \textbf{\bibinfo{volume}{73}},
  \bibinfo{pages}{35321} (\bibinfo{year}{2006}).

\bibitem[{\citenamefont{Perdew et~al.}(1996{\natexlab{b}})\citenamefont{Perdew,
  Burke, and Ernzerhof}}]{Perdew1996a}
\bibinfo{author}{\bibfnamefont{J.~P.} \bibnamefont{Perdew}},
  \bibinfo{author}{\bibfnamefont{K.}~\bibnamefont{Burke}}, \bibnamefont{and}
  \bibinfo{author}{\bibfnamefont{M.}~\bibnamefont{Ernzerhof}},
  \bibinfo{journal}{Phys. Rev. Lett.} \textbf{\bibinfo{volume}{77}},
  \bibinfo{pages}{3865} (\bibinfo{year}{1996}{\natexlab{b}}).

\bibitem[{\citenamefont{Perdew et~al.}(1997)\citenamefont{Perdew, Burke, and
  Ernzerhof}}]{Perdew1997a}
\bibinfo{author}{\bibfnamefont{J.~P.} \bibnamefont{Perdew}},
  \bibinfo{author}{\bibfnamefont{K.}~\bibnamefont{Burke}}, \bibnamefont{and}
  \bibinfo{author}{\bibfnamefont{M.}~\bibnamefont{Ernzerhof}},
  \bibinfo{journal}{Phys. Rev. Lett.} \textbf{\bibinfo{volume}{78}},
  \bibinfo{pages}{1396(E)} (\bibinfo{year}{1997}).

\bibitem[{\citenamefont{Perdew et~al.}(1998)\citenamefont{Perdew, Burke, Zupan,
  and Ernzerhof}}]{Perdew1998a}
\bibinfo{author}{\bibfnamefont{J.}~\bibnamefont{Perdew}},
  \bibinfo{author}{\bibfnamefont{K.}~\bibnamefont{Burke}},
  \bibinfo{author}{\bibfnamefont{A.}~\bibnamefont{Zupan}}, \bibnamefont{and}
  \bibinfo{author}{\bibfnamefont{M.}~\bibnamefont{Ernzerhof}},
  \bibinfo{journal}{J. Chem. Phys.} \textbf{\bibinfo{volume}{108}},
  \bibinfo{pages}{1522} (\bibinfo{year}{1998}).

\bibitem[{\citenamefont{{Delle Site} et~al.}(2002)\citenamefont{{Delle Site},
  Abrams, Alavi, and Kremer}}]{Dellesite2002a}
\bibinfo{author}{\bibfnamefont{L.}~\bibnamefont{{Delle Site}}},
  \bibinfo{author}{\bibfnamefont{C.}~\bibnamefont{Abrams}},
  \bibinfo{author}{\bibfnamefont{A.}~\bibnamefont{Alavi}}, \bibnamefont{and}
  \bibinfo{author}{\bibfnamefont{K.}~\bibnamefont{Kremer}},
  \bibinfo{journal}{Phys. Rev. Lett.} \textbf{\bibinfo{volume}{89}},
  \bibinfo{pages}{156103} (\bibinfo{year}{2002}).

\bibitem[{\citenamefont{Abrams et~al.}(2003)\citenamefont{Abrams, {Delle Site},
  and Kremer}}]{Abrams2003a}
\bibinfo{author}{\bibfnamefont{C.~F.} \bibnamefont{Abrams}},
  \bibinfo{author}{\bibfnamefont{L.}~\bibnamefont{{Delle Site}}},
  \bibnamefont{and} \bibinfo{author}{\bibfnamefont{K.}~\bibnamefont{Kremer}},
  \bibinfo{journal}{Phys. Rev. E} \textbf{\bibinfo{volume}{67}},
  \bibinfo{pages}{21807} (\bibinfo{year}{2003}).

\bibitem[{\citenamefont{Neaton et~al.}(2006)\citenamefont{Neaton, Hybertsen,
  and Louie}}]{Neaton2006a}
\bibinfo{author}{\bibfnamefont{J.}~\bibnamefont{Neaton}},
  \bibinfo{author}{\bibfnamefont{M.~S.} \bibnamefont{Hybertsen}},
  \bibnamefont{and} \bibinfo{author}{\bibfnamefont{S.~G.} \bibnamefont{Louie}},
  \bibinfo{journal}{Phys. Rev. Lett.} \textbf{\bibinfo{volume}{97}},
  \bibinfo{pages}{216405} (\bibinfo{year}{2006}).

\end{thebibliography}

\end{document}